\documentclass[aps,showpacs,floatfix]{revtex4}
\usepackage{graphicx}
\usepackage{graphics}
\usepackage{psfrag}
\usepackage{color}      
\usepackage{subfigure}

\newcommand{\bs}{\mathbf {s}}
\newcommand{\br}{\mathbf {r}}
\newcommand{\bv}{\mathbf {v}}

\newcommand{\bom}{{\mbox{\boldmath $\omega$}}}

\begin{document}

%+++++++++++++++++++++++++++++++++++++++++++++++++++++++++
\title{Spectrum of turbulent Kelvin-waves cascade in superfluid helium}
%+++++++++++++++++++++++++++++++++++++++++++++++++++++++++
\author{Andrew W. Baggaley$^1$}
\email{a.w.baggaley@ncl.ac.uk}
\author{Carlo F. Barenghi$^1$}
\email{c.f.barenghi@ncl.ac.uk}

\affiliation{$^1$ School of Mathematics and Statistics, 
Newcastle University, Newcastle upon Tyne, NE1 7RU, England, UK}

\date{\today}
%++++++++++++++++++++++++++++++++++++++++++++++++++++++++
\begin{abstract}
To explain the observed decay of superfluid turbulence at
very low temperature, it has been proposed that a
cascade of Kelvin waves (analogous to the classical Kolmogorov
cascade) transfers kinetic energy to length scales which
are small enough that sound can be radiated away. 
We report results of numerical simulations of the 
interaction of quantized vortex filaments. We observe the
development of the Kelvin-waves cascade, and compute the statistics
of the curvature, the amplitude spectrum 
(which we compare with competing theories) and
the fractal dimension.
\end{abstract}
%++++++++++++++++++++++++++++++++++++++++++++++++++++++++++++++++++++++++++++++++++
\maketitle

\section{Motivation}\label{I}

Quantum turbulence \cite{Donnelly,BDV,BS} consists of a disordered tangle of 
reconnecting superfluid vortex filaments. Because of quantum mechanical
constraints on the rotational motion, these
filaments carry the same quantized (fixed) circulation $\Gamma$
and are very thin: the (fixed) vortex core radius, $a_0$, 
is much smaller than the average distance between the filaments, $\ell$.
Quantum turbulence is easily created by agitating superfluid helium ($^4$He)
with propellers \cite{propeller}, grids \cite{grid}, 
forks \cite{fork} or wires \cite{wire}, by applying a 
heat flow \cite{Vinen-counterflow,Tough-counterflow,
Paoletti-counterflow,Barenghi-Skrbek,Skrbek-2003}
or by injecting a stream of ions \cite{Golov}.  
It is also studied in superfluid $^3$He-B \cite{Helsinki,Lancaster} and, 
more recently, in atomic Bose-Einstein condensates\cite{Bagnato,White}.  

If the temperature is relatively large (more than
$1~\rm K$ in $^4$He), the turbulent kinetic energy contained in the
superfluid vortices is transferred by the mutual friction \cite{friction}
into the viscous gas of thermal excitations (the normal fluid) 
and then decays into heat; therefore a constant supply of energy 
(continuous stirring for example) is needed
to maintain the intensity of the turbulence.
If the temperature is relatively small (less than $1~\rm K$
in $^4$He), the normal fluid is negligible, but, despite the absence of
viscous dissipation,  the turbulence still decays \cite{McClintock,Bradley}.
The Kelvin-waves cascade \cite{Svistunov,Kivotides-cascade,Vinen-cascade}
was proposed to explain this surprising effect.

A Kelvin wave is a rotating sinusoidal or helical displacement
of the core of vortex filaments away from its unperturbed position 
\cite{LordKelvin,JJThomson,Pocklington}. The dispersion
relation of a Kelvin wave of angular frequency
$\omega$ and wavenumber $k$ along a straight vortex is \cite{BDV-PoF}

\begin{equation}
\omega=\frac{\Gamma}{2 \pi a_0^2} 
\left( 1-\sqrt{1+\frac{K_0(k a_0)}{K_1(k a_0)}}
\right),
\label{eq:Kelvin}
\end{equation}

\noindent
where $a_0$ is the vortex core radius and
$K_n$ is the modified Bessel function of order $n$. In the 
long-wavelength approximation ($k a_0 \ll 1$), the angular frequency
reduces to
\begin{equation}
\omega \approx - \frac{\Gamma k^2}{4 \pi}  
\left( \ln{2/(k a_0)}-\gamma \right)
= - \frac{\Gamma k^2}{4 \pi}  
\left( \ln{1/(k a_0)}-0.116 \right)
\label{eq:dispersion}
\end{equation}
\noindent
where $\gamma=0.5772$ is Euler's constant; the negative sign in this
expression means that the Kelvin wave propagates in the
direction opposite to the orientation of the unperturbed vorticity.
Eq.~\ref{eq:Kelvin} was originally derived for a thin, hollow--cored
vortex in a perfect Euler fluid; a similar expression was obtained
for a vortex in a Bose--Einstein condensate \cite{Grant, Pitaevskii}.
The simplified dispersion
formula $\omega(k)=c k^2$, where $c$ is a constant, is often used in the 
Kelvin-waves cascade literature.

The Kelvin-waves cascade is the process in which 
the nonlinear interaction of Kelvin waves
creates waves of shorter and shorter
wavelength $\lambda=2 \pi/k$.  At high enough temperature
the mutual friction would quickly damp out the shorter Kelvin 
waves \cite{BDV-PoF}, but at low
temperatures the cascade proceeds unhindered, until $k$
is large enough that sound is efficiently radiated away (phonon
emission) by rapidly rotating 
vortices \cite{Vinen-sound,Leadbeater-sound,Parker-sound,Kozik-sound}.
There is thus an analogy between the Kelvin-waves cascade in
quantum turbulence (where the energy sink is acoustic) 
and the Kolmogorov cascade \cite{Frisch,Davidson}
in classical turbulence (where the energy sink
is viscous).

It is thought that in superfluid helium at low temperatures both
the Kolmogorov cascade and the Kelvin-waves cascades operate 
at the same time \cite{Vinen-Niemela}. Let $\Lambda=L/V$ be the vortex line
density (vortex length $L$ in the volume $V$) of a
homogeneous isotropic vortex tangle, and $\ell \approx \Lambda^{-1/2}$
be the average intervortex spacing.
The Kolmogorov cascade (which in a classical fluid is often described
as the break up of eddies \cite{Goto}) involves rearrangements or
partial polarisation\cite{Hulton} of vortex lines over length scales of
wavenumbers $k<<2 \pi/\ell$.
The Kelvin-waves cascade occurs on individual vortex filaments 
for wavenumbers $k>>2 \pi/\ell$. Recent work has raised the possibility of
the existence of a  bottleneck 
\cite{Lvov-Nazarenko-Rudenko-PRB-2007,Lvov-Nazarenko-Rudenko-JLTP-2008}
between the two cascades, 
for which energy would pile up at some length scale of the order of $\ell$.

Direct evidence of the Kolmogorov cascade in
superfluid helium was provided by the experiment of 
Maurer and Tabeling \cite{Tabeling}. 
Numerical evidence is also convincing: the
Kolmogorov $E_k \sim k^{-5/3}$ energy spectrum has been seen in calculations
performed using both the Schwarz filament model \cite{Araki}
and the Gross--Pitaevskii equation 
\cite{Nore,Kobayashi-Tsubota,Yepez}
which describes a Bose-Einstein condensate.  
On the contrary, the Kelvin-waves cascade is less understood; 
there is no direct observation of it, and the theory is mired in
a controversy between approaches proposed by
L'vov, Nazarenko and collaborators 
\cite{Lvov-Nazarenko-Volovik-JETP-2004,
Lvov-Nazarenko-Skrbek-JLTP-2006,
Nazarenko-JETP-2006,
Lvov-Nazarenko-Rudenko-PRB-2007,Lvov-Nazarenko-Rudenko-JLTP-2008,
Proment-Nazarenko-Onorato-PRA-2009,
Boffetta-Laurie-Nazarenko-JLTP-2009,
Lvov-Nazarenko-arXiv-2010}
on one hand,
and by Kozik and Svistunov
\cite{Svistunov,Kozik-Svistunov-PRL-2004,Kozik-Svistunov-PRL-2005,
Kozik-sound,Kozik-Svistunov-PRL-2008,
Kozik-Svistunov-PRB-2008,Kozik-Svistunov-JLTP-2009} on the other.
The two theories predict similar Kelvin-waves amplitude spectra,
but differ on matters of principles
and in some important details. 

Lacking direct experimental evidence of the 
Kelvin-waves cascade and its properties, experimentalists currently attempt to 
identify indirect but observable effects of the cascade and of  
any bottleneck;
for example they study the temperature dependence of 
the observed effective kinematic viscosity \cite{Chagovets,Golov-PRL-2007}.
The results are still inconclusive.

The aim of our work at this stage
is not to review or compare rival theories, but to add information.
We present results of numerical calculations of the Kelvin-waves cascade.
The few published calculations differ in details (e.g.
whether the initial vortex is curved or straight, whether the cascade
is forced continuously or results from an initial
condition - either an artificially imposed cusp or a dynamically
resolved vortex reconnection). 
More information is therefore necessary to learn the main physical
features of the turbulent Kelvin-waves cascade and to
assess  which results are robust.
In particular we shall focus the attention to the
magnitude and spectrum of the Kelvin waves,
the statistics of the vortex curvature and the fractal dimension.

Finally we note that, in presenting our results, we shall pay
particular attention to how well energy is conserved in a given
numerical approximations. In the
absence of mutual friction, 
vortex evolution in between vortex reconnection events and without
a turbulent cascade which radiates sound, energy should be conserved;
yet previously published numerical calculations of the evolution of 
superfluid vortex filaments do not
report how well the conservation law is satisfied.

\section{Model}\label{II}

Hereafter we use parameters which refer to superfluid $^4$He: circulation 
$\Gamma=9.97 \times 10^{-4}~\rm cm^2/s$ and vortex core radius
$a_0\approx 10^{-8}~\rm cm$, but our results can be generalised to 
turbulence in low temperature $^3$He-B. 
Following Schwarz \cite{Schwarz}, we describe
vortex filaments as space curves ${\bs}={\bs}(\xi,t)$ 
where $\xi$ is arc length
and $t$ is time. In the absence of mutual friction and of any externally
applied superflow, the self-induced
velocity of a superfluid filament at the point ${\bs}$ is
given by the Biot-Savart law

\begin{equation}
\frac{d\bs}{dt}=-\frac{\Gamma}{4 \pi} \oint_{\cal L} \frac{(\bs-\br) }
{\vert \bs - \br \vert^3}
\times {\bf d}\br.
\label{eq:BS}
\end{equation}

\noindent
The line integral extends over the entire vortex configuration $\cal L$,
which is discretised into a large number of points $\bs_i$
($i=1,\cdots N$).
The singularity at $\bs=\br$ is removed in a standard way 
by considering local and non-local contributions to the integral. 
If $\bs_i$ is the position of the $i^{\rm th}$ discretization point
along the vortex line, Eq.~\ref{eq:BS} becomes \cite{Schwarz2}

\begin{equation}
\frac{d\bs_i}{dt}=
\frac{\Gamma}{4\pi} \ln \left(\frac{\sqrt{\ell_i \ell_{i+1}}}{a}\right)\bs_i' \times \bs_i'' 
-\frac{\Gamma}{4 \pi} \oint_{\cal L'} \frac{(\bs_i-\br) }
{\vert \bs_i - \br \vert^3}
\times {\bf d}\br.
\label{eq:BS_sing}
\end{equation}

\noindent
Here $\ell_i$ and $\ell_{i+1}$ are the arclengths of the curve 
between points $\bs_{i-1}$ and $\bs_i$ and between $\bs_i$ and $\bs_{i+1}$, 
and $\cal L'$ is the original vortex line without the section 
between $\bs_{i-1}$ 
and $\bs_{i+1}$. For simplicity we assume that the cutoff parameter is
$a\approx a_0$.  Other de-singularization techniques are also used in the
fluid dynamics literature\cite{Leonard1985,Siggia}.

The number of discretization points, $N$, changes with time.
As the simulation progresses, new discretization points are introduced 
to maintain the resolution along the vortex filament.
If the separation between two points, $\bs_{i}$ and $\bs_{i+1}$, 
becomes greater than some threshold $\delta$ (which we call the 
minimum resolution), we introduce a new point at position $\bs_{i'}$ given by
\begin{equation}
\bs_{i'}=\frac{1}{2}(\bs_i+\bs_{i+1})+\left( \sqrt{R^2_{i'}
-\frac{1}{4}\ell_{i+1}^2}-R_{i'} \right)\frac{\bs_{i'}^{''}}{|\bs_{i'}^{''}|},
\end{equation}

\noindent
where $R_{i'}=|\bs_{i'}^{''}|^{-1}$. In this way 
$\bs^{''}_{i'}=(\bs^{''}_i+\bs^{''}_{i+1})/2$,
that is to say the insertion of new points preserves the curvature.
Points are removed if their separation is less than $\delta/2$, 
to ensure that our shortest length-scale is fixed.
The above algorithms ensure that the separation of vortex points along 
a filament lies between $\delta/2$ and $\delta$.

For accuracy, we approximate all spatial derivatives using 
$4^{\rm th}$ order finite difference schemes which account 
for varying mesh sizes along the vortex filaments \cite{FiniteDiff}.
Let $\bs_i$ be the $i^{\rm th}$ point on the vortex filament; the two
points behind have positions
$\bs_{i-2}$ and $\bs_{i-1}$, and the two points in front have positions 
$\bs_{i+1}$ and $\bs_{i+2}$.
We denote $\ell_{i-1}=|\bs_{i-1}-\bs_{i-2}|$, $\ell_i=|\bs_{i}-\bs_{i-1}|$, 
$\ell_{i+1}=|\bs_{i+1}-\bs_{i}|$, and $\ell_{i+2}=|\bs_{i+2}-\bs_{i+1}|$.
We can then write
\begin{equation}
  \bs_i'=A_{1i}\bs_{i-2}+B_{1i}\bs_{i-1}+C_{1i}\bs_{i}+D_{1i}\bs_{i+1}+E_{1i}\bs_{i+2},
\end{equation}
where the coefficients $A_{1i}$, $B_{1i}$, $C_{1i}$, $D_{1i}$ and $E_{1i}$ are given by,
\begin{equation}
  A_{1i}= \frac{\ell_{i}\ell_{i+1}^2+\ell_{i}\ell_{i+1}\ell_{i+2}}{\ell_{i-1}(\ell_{i-1}+\ell_{i})(\ell_{i-1}+\ell_{i}+\ell_{i+1})(\ell_{i-1}+\ell_{i}+\ell_{i+1}+\ell_{i+2})}
\end{equation}
\begin{equation}
  B_{1i}=\frac{-\ell_{i-1}\ell_{i+1}^2-\ell_{i}\ell_{i+1}^2-\ell_{i-1}\ell_{i+1}\ell_{i+2}-\ell_{i}\ell_{i+1}\ell_{i+2}}{\ell_{i-1}\ell_{i}(\ell_{i}+\ell_{i+1})(\ell_{i}+\ell_{i+1}+\ell_{i+2})}
\end{equation}
\begin{equation}
  C_{1i}=-(A_{1i}+B_{1i}+D_{1i}+E_{1i})
\end{equation}
\begin{equation}
  D_{1i}=\frac{\ell_{i-1}\ell_{i}\ell_{i+1}+\ell_{i}^2\ell_{i+1}+\ell_{i-1}\ell_{i}\ell_{i+2}+\ell_{i}^2\ell_{i+2}}{\ell_{i+1}\ell_{i+2}(\ell_{i}+\ell_{i+1})(\ell_{i-1}+\ell_{i}+\ell_{i+1})}
\end{equation}
\begin{equation}
  E_{1i}= \frac{-\ell_{i+1}\ell_{i}^2-\ell_{i-1}\ell_{i}\ell_{i+1}}{\ell_{i+2}(\ell_{i+1}+\ell_{i+2})(\ell_{i}+\ell_{i+1}+\ell_{i+2})(\ell_{i-1}+\ell_{i}+\ell_{i+1}+\ell_{i+2})}
\end{equation}
In a similar manner we can write,
\begin{equation}
  \bs_i''=A_{2i}\bs_{i-2}+B_{2i}\bs_{i-1}+C_{2i}\bs_{i}+D_{2i}\bs_{i+1}+E_{2i}\bs_{i+2},
\end{equation}
where the coefficients $A_{2i}$, $B_{2i}$, $C_{2i}$, $D_{2i}$ and $E_{2i}$ are given by,
\begin{equation}
  A_{2i}= \frac{2[-2\ell_{i}\ell_{i+1}+\ell_{i+1}^2-\ell_{i}\ell_{i+2}+\ell_{i+1}\ell_{i+2}]}{\ell_{i-1}(\ell_{i-1}+\ell_{i})(\ell_{i-1}+\ell_{i}+\ell_{i+1})(\ell_{i-1}+\ell_{i}+\ell_{i+1}+\ell_{i+2})}
\end{equation}
\begin{equation}
  B_{2i}= \frac{2[2\ell_{i-1}\ell_{i+1}+2\ell_{i}\ell_{i+1}-\ell_{i+1}^2+\ell_{i-1}\ell_{i+2}+\ell_{i}\ell_{i+2}-\ell_{i+1}\ell_{i+2}]}{\ell_{i-1}\ell_{i}(\ell_{i}+\ell_{i+1})(\ell_{i}+\ell_{i+1}+\ell_{i+2})}
\end{equation}
\begin{equation}
  C_{2i}=-(A_{2i}+B_{2i}+D_{2i}+E_{2i})
\end{equation}
\begin{equation}
  D_{2i}= \frac{2[-\ell_{i-1}\ell_{i}-\ell_{i}^2+\ell_{i-1}\ell_{i+1}+2\ell_{i}\ell_{i+1}+\ell_{i-1}\ell_{i+2}+2\ell_{i}\ell_{i+2}]}{\ell_{i+1}\ell_{i+2}(\ell_{i}+\ell_{i+1})(\ell_{i-1}+\ell_{i}+\ell_{i+1})}
\end{equation}
\begin{equation}
  E_{1i}= \frac{2[\ell_{i-1}\ell_{i}+\ell_{i}^2-\ell_{i-1}\ell_{i+1}-2\ell_{i}\ell_{i+1}]}{\ell_{i+2}(\ell_{i+1}+\ell_{i+2})(\ell_{i}+\ell_{i+1}+\ell_{i+2})(\ell_{i-1}+\ell_{i}+\ell_{i+1}+\ell_{i+2})}
\end{equation}
Note that if we set $\ell_{i-1}=\ell_i=\ell_{i+1}=\ell_{i+2}=h$, then the above expressions reduce to familiar 
finite-difference schemes for a uniform mesh:
\begin{equation}
\bs_i'=\frac{1}{12h}(\bs_{i-2}-8\bs_{i-1}+8\bs_{i+1}-\bs_{i+2})+\mathcal{O}(h^4),
\end{equation}
\begin{equation}
\bs_i''=\frac{1}{12h^2}(-\bs_{i-2}+16\bs_{i-1}-30\bs_{i}+16\bs_{i+1}-\bs_{i+2})+\mathcal{O}(h^4).
\end{equation}

The Biot-Savart law describes an incompressible
flow. In order to model energy dissipation at large $k$
(phonon emission), we also remove points if the local 
wavelength is smaller than a specified value.
Since the smallest length-scale is $\delta/2$, the maximum curvature is 
of the order of $2/\delta$. 
At each time step we compute the local curvature 
$C(\xi)=\vert \bs'' \vert=\vert d^2\bs/d\xi^2 \vert$ 
at each discretization point $\bs_j$ ($j=1,\cdots N$)  
If, at some location along a vortex filament, the local curvature exceeds 
the critical value $1.9/\delta~\mathrm{cm}^{-1}$
($95 \%$ of the maximum value $2/\delta$), points are
removed and the filament is smoothed locally.
This leads to a loss of line length, hence a loss of energy.

It must be stressed that
we are not interested in the phonon emission per se, but rather in the 
consequence of the presence of an energy sink on the Kelvin waves
cascade. The actual numerical value of $\delta$ is chosen for
numerical convenience and is orders of magnitude larger than the 
microscopic lengthscales which are physically relevant to phonon emission 
and vortex reconnections. Therefore our model contains (and controls) 
a mechanism of dissipation, and is different in scope from vortex methods
dedicated to perfect Euler dynamics.
In the next section we describe results
obtained with two different spatial resolutions: 
$\delta=10^{-3}~\rm cm$ (high resolution) and
$\delta=5 \times 10^{-3}~\rm cm$ (low resolution). We shall see that
our results are independent of the resolution used.

Time evolution is based on a $3^{\rm rd}$ order Adams--Bashforth scheme.
Consider an evolution equation of the form $d\bs/dt=\bv$. We time-step
the equation according to
\begin{equation}
  \bs_{i}^{n+1}=\bs_{i}^{n}+\frac{\Delta t}{12}(23\bv_{i}^{n}-16\bv_{i}^{n-1}+5\bv_{i}^{n-2})+\mathcal{O}(\Delta t^4)
\end{equation}
\noindent
where $\Delta t$ is the time step and the superscript $n$ refers to the time $t_n=n\Delta t$ $(n=0,1,2,\cdots)$.
Lower order schemes are used for the initialisation, when older velocity values are not available. 
The time step $\Delta t$ is chosen so 
that the period $\tau_{min}=2 \pi/\omega_{max}$ of the fastest Kelvin wave
(which has wavenumber 
$k_{max}=2 \pi/\lambda_{min} \approx \pi/\delta$) 
contains at least 25 time steps; we have
verified that reducing our time step further gives the same 
results.

If two discretization points become closer to each other than the local
discretization distance, our numerical algorithm reconnects the 
two filaments \cite{Koplik}
subject to the criteria that the total length (as proxy for energy)
decreases \cite{Leadbeater-recon}. In this way the same quantity $\delta$
controls both energy dissipation mechanisms in our model - sound emission and
reconnections. Self-reconnections (which can arise 
if a vortex filament has twisted and coiled by a large amount) are treated
in the same way.
Since reconnections involve only anti-parallel filaments, prior
to reconnection we form local (unit) tangent vectors $\bs'=d{\bs}/d\xi$,
where $\xi$ is arc length, and,  
using the inner product, we check that the two filaments are not parallel.
Finally, We remove any small loops with less than three discretization points.

During the evolution we monitor the length $L$ 
and the kinetic energy $E$ (in unit of
the density) of the vortex configuration; the latter is

\begin{equation}
E=\frac{1}{2} \int_V \bv^2 dV.
\label{eq:E0}
\end{equation}

\noindent
If the calculation is performed in a periodic box, the vortices can
extend across the boundaries; the only way to compute the energy integral 
is to discretise the integrand on a three--dimensional spatial grid, computing
the velocity at each grid point; unfortunately this procedure converges
too slowly to be practical. However, if
the calculation is performed in an infinite
volume, the vortex filaments are closed loops. Assuming that
the velocity field decays to zero at infinity, the energy integral can be
transformed into a line integral \cite{Saffman}

\begin{equation}
E=\frac{1}{2} \int_V \bv^2 dV
=\int_V \bv \cdot \br \times \bom dV
=\Gamma \oint_{\cal L} \bv \cdot \br \times \bs' d\xi,
\label{eq:E}
\end{equation}

\noindent
which converges much better. In writing Eq.~\ref{eq:E}
we have used the fact that the vorticity $\bom=\nabla \times \bv$ is 
concentrated on the vortex filaments ($\bom dV=\Gamma \bs' d\xi$).

In the absence of mutual friction,
during Biot-Savart evolution (but not during vortex reconnections, and not
while the Kelvin-waves cascade is in progress with the associated
sound emission which we model), 
the kinetic energy $E$ is conserved, hence changes in $E$ measure
the accuracy of the numerical method. Our current method conserves
energy about a factor of ten better than our preliminary
unpublished calculations. The energy-conserving properties of
previous calculations in the superfluid literature
is unknown.

Unlike the energy $E$, the vortex length $L$ is not conserved. 
However, if the Biot-Savart law is replaced by the Local
Induction Approximation (LIA), the length is conserved. The LIA
was used in the original work of Schwarz \cite{Schwarz} and by others
to reduce the computational cost. We do not use the LIA
in our calculations.

A quantity which is particularly interesting is the normalized
histogram (probability density function, of PDF for short) of the
curvature $C(\xi)$, sampled on the discretization points: 
we call this quantity ${\rm PDF}(C)$.

\section{Results}\label{III}
\subsection{Cascading and non-cascading vortex configurations}\label{III.A}

\begin{figure}
\begin{minipage}{8cm}
\includegraphics[width=1.15\textwidth]{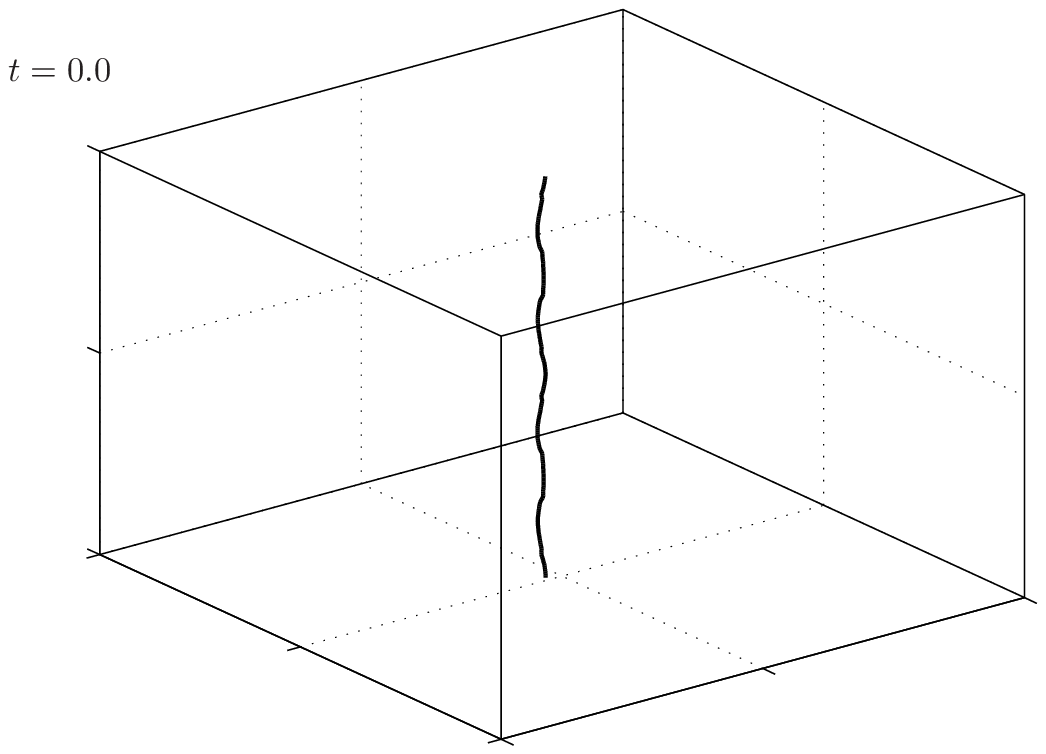}
\end{minipage}
\caption{
Single vortex line  
initially perturbed by
two Kelvin waves of wavenumbers $k=3$ and $k=4$ (where $k=2 \pi/\lambda$)
and amplitudes $A=3\delta$ and $A=1.2 \delta$, where $\delta= 0.001 \rm cm$.
The vortex line is contained in a cubic periodic box of size $D=0.1~\rm cm$. 
This initial condition does not evolve into a Kelvin-waves cascade.
}
\label{fig:1}
\end{figure}

\begin{figure}
\begin{minipage}{8cm}
\includegraphics[width=0.95\textwidth]{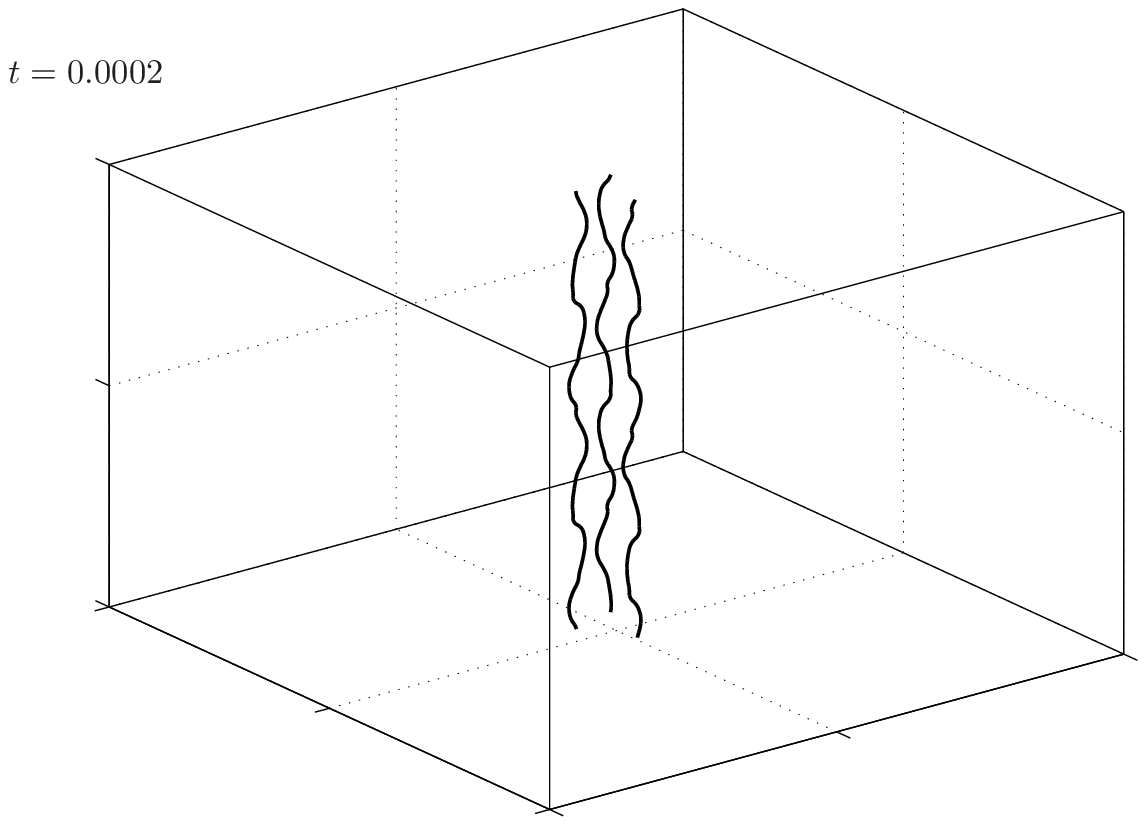}
\end{minipage}
\begin{minipage}{8cm}
\includegraphics[width=0.95\textwidth]{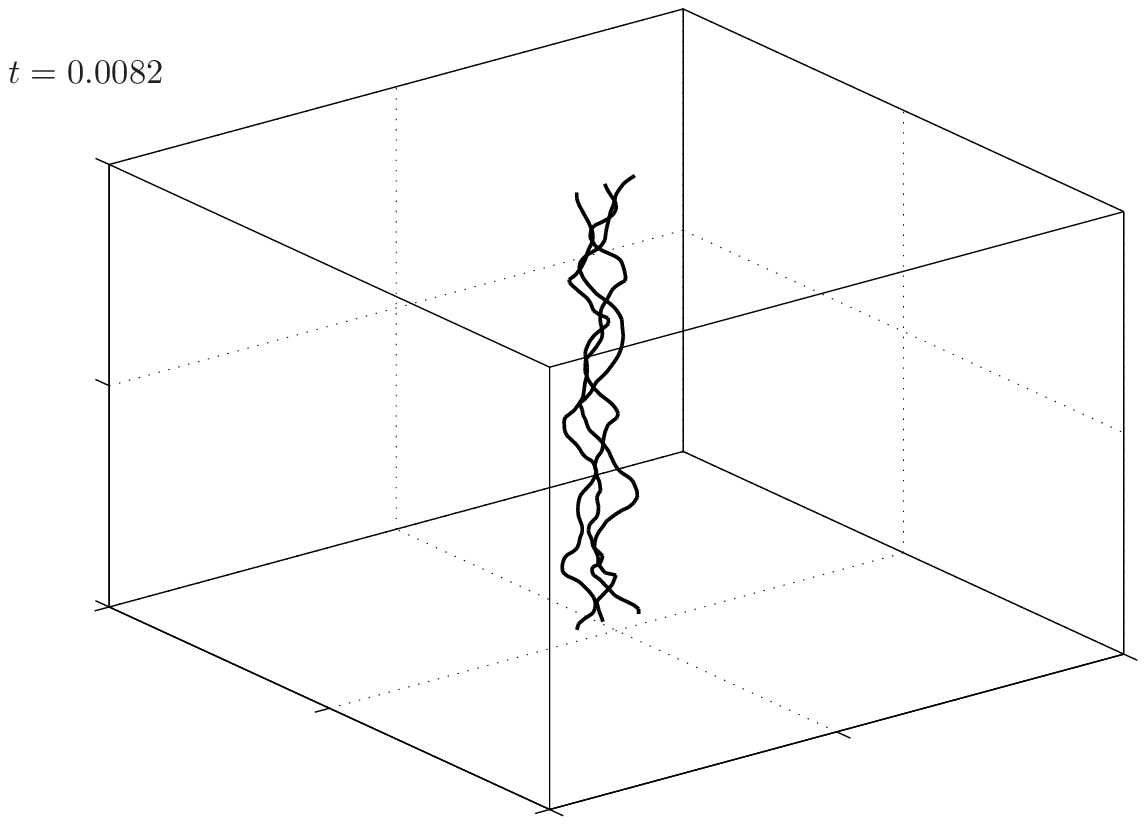}
\end{minipage}
\caption{
Time evolution of three parallel vortex lines in a cubic periodic
box of size $D=0.1~\rm cm$. Initially, at $t=0$ (left),
each vortex line is
perturbed by two Kelvin waves of wavenumbers $k=3$ and $k=4$ 
and amplitudes $A=3\delta$ and $A=1.2 \delta$, as in Fig.~\ref{fig:1}.
The waves are offset by random phases. This initial condition
triggers a Kelvin-waves cascade, as shown by the increasing
wiggliness at time $t=0.0082~\rm s$ (right).
}
\label{fig:2}
\end{figure}

In the first set of numerical experiments we consider a single vortex filament
set across a periodic box as in Fig.~\ref{fig:1}. 
The initial shape of the vortex is not straight, but is
perturbed by two sinusoidal Kelvin waves of amplitude $A$ and different
wavelengths $\lambda$.  We compute the time evolution
and observe that that this initial condition does not yield a cascade 
to shorter waves.
However, if we start with three parallel vortex filaments
(each filament with two Kelvin waves), as shown in Fig.~\ref{fig:2}(left),
we notice that, as the
vortices rotate around each other, they become more wiggly, 
see Fig.~\ref{fig:2}(right).  

This result suggests that,
besides vortex reconnection events (as shown in 
Ref.~\cite{Kivotides-cascade}), the Kelvin-waves cascade can be initiated,
more in general, by
the interaction of nearby curved vortices via the velocity field which
they induce on each other.
The result is also consistent with  
wave-turbulence theory \cite{Jason}, which states that, for one--dimensional
waves obeying a dispersion law of the form $\omega(k)=c k^2$, a four-waves
process is not sufficient to generate a cascade. According to the theory,
the minimal initial condition for the cascade consists of three
waves which satisfy the resonant condition that $k_1+k_2+k_3=k_4+k_5+k_6$
and
$\omega(k_1)+\omega(k_2)+\omega(k_3)=\omega(k_4)+\omega(k_5)+\omega(k_6)$,
where indices $1,2,3$ and $4,5,6$ refer to incoming and outgoing waves
respectively. The initial condition shown in Fig.~\ref{fig:2}(left) contains
more waves than the minimal number of waves (if it contained only the minimal 
number of waves we would 
need to seek the exact resonant condition, which is not our purpose 
at this stage).

In order to pursue this investigation more quantitatively,
we want to monitor the kinetic energy of the vortex configuration. 
Unfortunately, as we have mentioned in Section~\ref{II},
if the vortex configurations is set in a periodic box,
it is not possible to accurately determine its energy. 
Hereafter, in order to use Eq.~\ref{eq:E}, 
we consider only closed vortex loops contained in an infinite volume.

\begin{figure}
\begin{minipage}{8cm}
\includegraphics[width=0.95\textwidth]{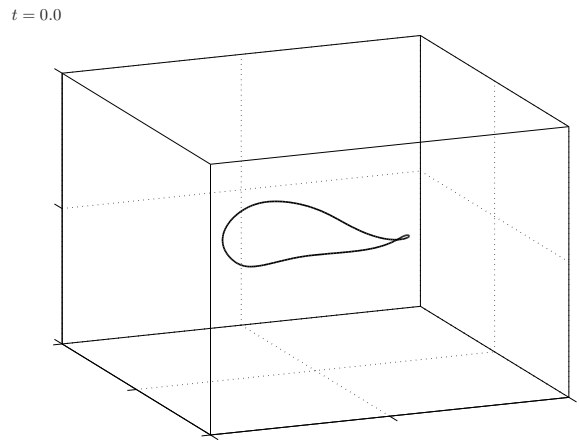}
\end{minipage}
\caption{
Single vortex ring of radius $R=0.1~\rm cm$ perturbed by two Kelvin waves
of mode numbers $m=2$ and $m=5$ (where $\lambda=2 \pi R/m$)
and amplitudes $A=5\delta$ and $A=1.4\delta$, where $\delta=0.001~\rm cm$.
The vortex ring is contained in an infinite volume and the box
(of size $D=0.1 \rm cm$) is for visualization only. 
This initial condition does not evolve into a Kelvin-waves cascade.
}
\label{fig:3}
\end{figure}

\begin{figure}

\begin{minipage}{11cm}
\begin{center}
\includegraphics[width=0.7\textwidth]{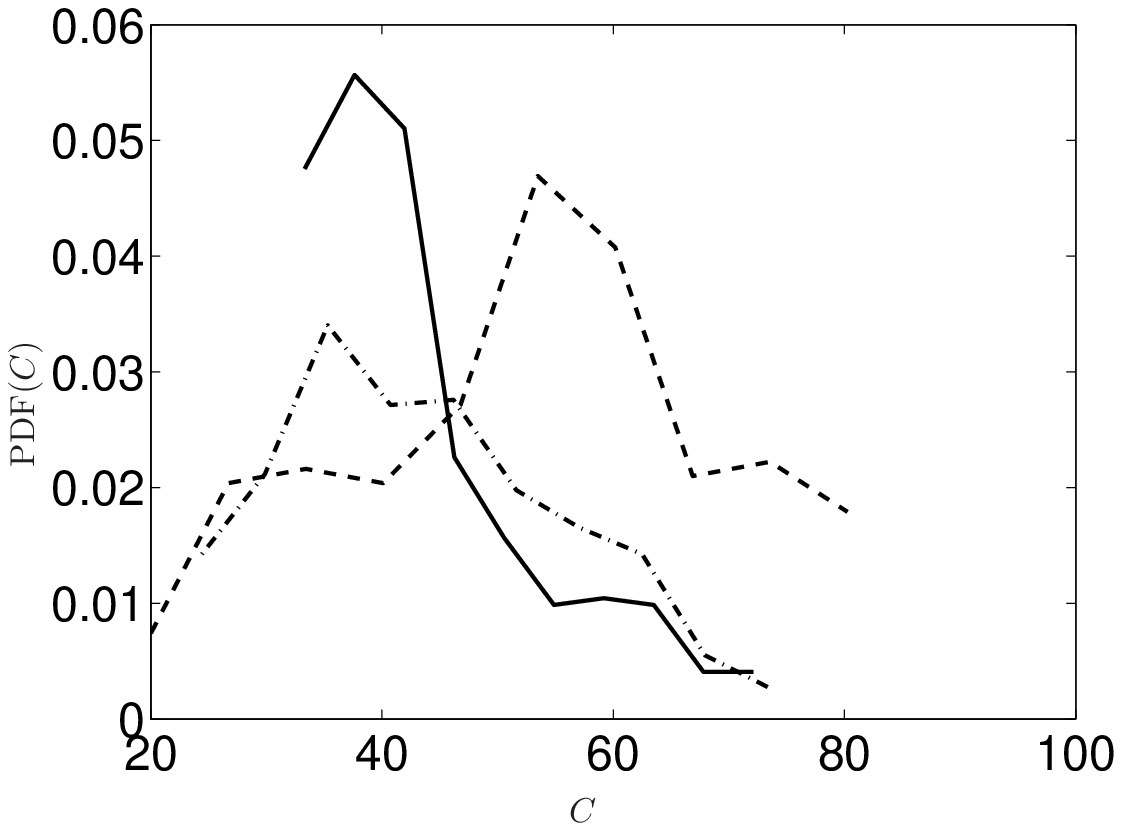}
\end{center}
\end{minipage}
\begin{minipage}{10cm}
\begin{center}
\includegraphics[width=1.25\textwidth]{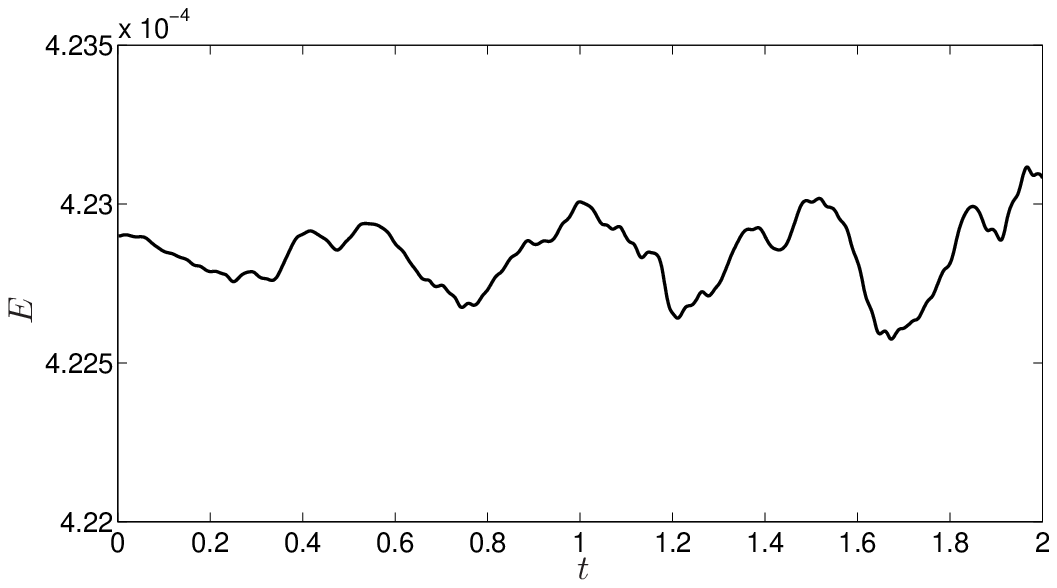}
\end{center}
\end{minipage}
\caption{
Evolution of curvature and energy corresponding to the non-cascading initial
condition shown in Fig.~\ref{fig:3}.
Top:
${\rm PDF}(C)$ vs $C~(\rm cm^{-1})$ at different times.
Solid line: at $t=0$; dashed line: at $t=0.5~\rm s$;
dot-dashed line: at $t=1~\rm s$.
Note that the peak of the PDF does not move significantly to
larger values of $C$ with time $t$.
Bottom:
Energy $E~(\rm cm^2/s^2)$ vs time $t~(\rm s)$.
}  
\label{fig:4}
\end{figure}

In a second set of numerical experiments we start with
a single vortex ring in an infinite volume; the ring is
seeded with two Kelvin waves,
as shown in Fig.~\ref{fig:3}. During the time evolution we observe
that the waves interact, but there is no cascade, consistently with

the evolution of the vortex line of Fig.~\ref{fig:1}. Indeed, if we
examine the PDF of the curvature, we notice that ${\rm PDF}(C)$
does not evolve in time to larger values of $C$,
as shown in Fig.~\ref{fig:4}(top).
Unlike the vortex line in the periodic box of Fig.~\ref{fig:1}, now
we can monitor how well the kinetic energy is conserved in our
numerical calculations.
Fig.~\ref{fig:4}(bottom) shows that the
relative change of energy during the time
evolution of this non-cascading vortex configuration of interacting
Kelvin waves
is only $\Delta E/E \approx \pm 0.06$ percent. This value describes
how well we can integrate Euler dynamics.

\subsection{Curvature and energy}\label{III.B}

\begin{figure}
\begin{minipage}{8cm}
\includegraphics[width=0.95\textwidth]{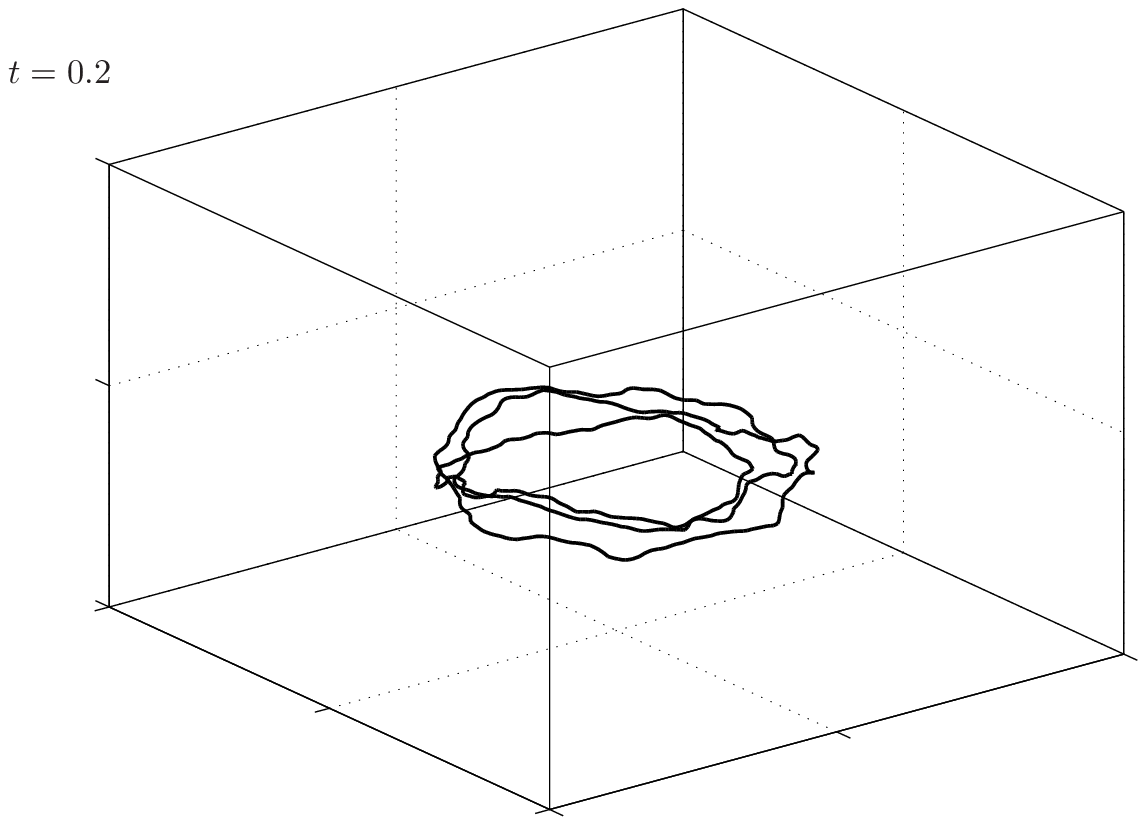}
\end{minipage}
\begin{minipage}{8cm}
\includegraphics[width=0.95\textwidth]{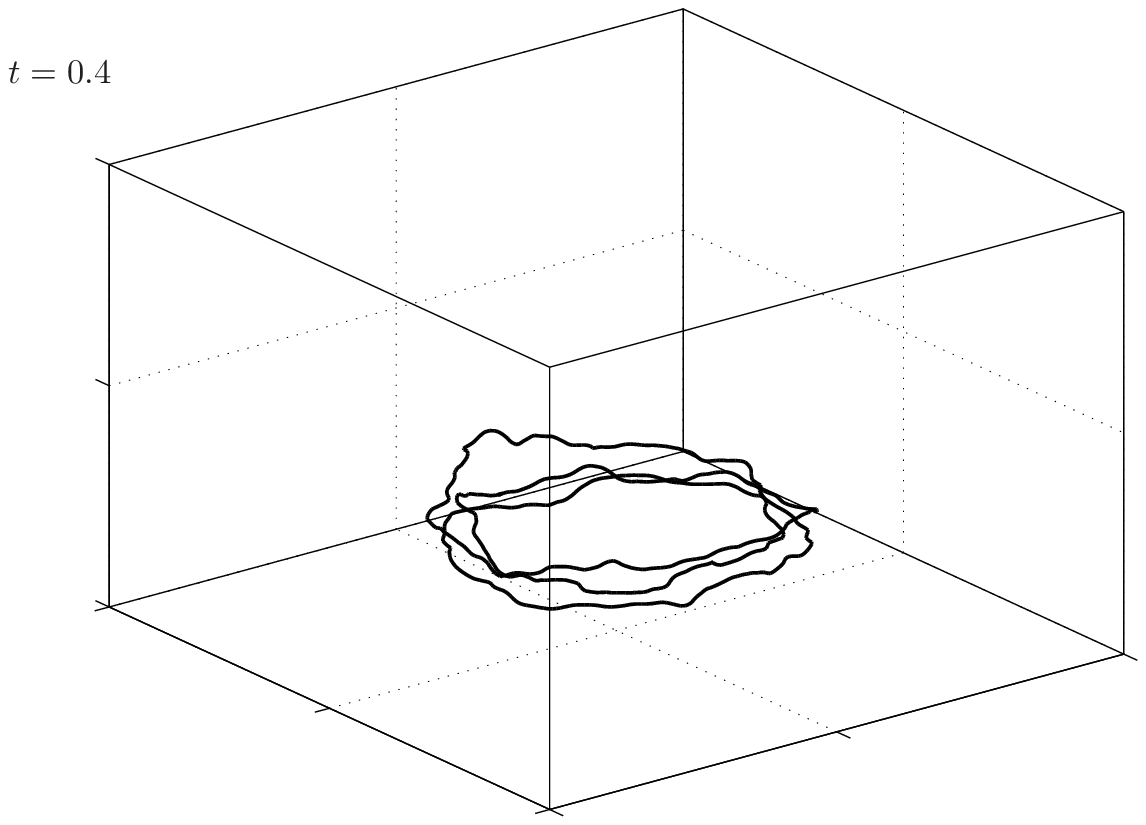}
\end{minipage}
\caption{
Cascading vortex configuration.
Time evolution of three vortex rings of radii 
$R_1=R_2=R_3=0.16~\rm cm$. Each ring has been perturbed by
two Kelvin waves. Two rings have waves with mode numbers $m=3$ and $m=4$
and amplitudes $A=5\delta$ and $A=1.4\delta$; one ring has waves
with $m=2$ and $m=3$ and $A=5\delta$ and $A=1.4\delta$.
As in Fig.~\ref{fig:2}, this initial condition triggers
a Kelvin-waves cascade (note the increasing wiggliness of the
filaments). The calculation is performed in an infinite volume: 
the box shown (for visualisation purpose) is a cube with sides 
of length $0.1 \rm cm$.
Top left: at time $t=0~\rm s$;
Top right: at $t=0.1~\rm s$;
Bottom left: at $t=0.2~\rm s$;
Bottom right: at $t=0.4~\rm s$.
}
\label{fig:5}
\end{figure}

Proceeding in analogy with Fig.~\ref{fig:2}, we now consider a third
set of numerical experiments with
three vortex rings oriented in the same direction as shown in the
first snapshot of
Fig.~\ref{fig:5}; each ring is
initially perturbed by two random Kelvin 
waves. This set-up contains
enough waves to trigger the Kelvin-waves cascade
and allows us to carefully monitor the energy.

The initial total number of
discretization points is $N_0=600$; at the end of the calculation
at time $t=2~\rm s$ we have $N=773$.
The minimum distance between discretization points is set to 
$\delta=5 \times 10^{-4}~\rm cm$.
All three rings are oriented in the same way, hence they travel in the
same (negative $z$) direction, leapfrogging around each other.
This simple vortex bundle configuration evolves into a Kelvin-waves cascade:
Fig.~\ref{fig:5} shows that during
the evolution the wiggliness of the vortices increases. This is in
agreement with the straight vortices of Fig.~\ref{fig:2}. 
Since the wiggly vortices of Fig.~\ref{fig:5}  are relatively
close to each other, there are many reconnections up to time $t=0.1~\rm s$,
less reconnections between $0.1 <t<0.5~\rm s$, and no reconnection
afterwards. The reconnections do not appear to affect the vortex
configuration in a significant way: the vortex bundle seems structurally
robust.

\begin{figure}
\begin{minipage}{12cm}
\includegraphics[width=0.95\textwidth]{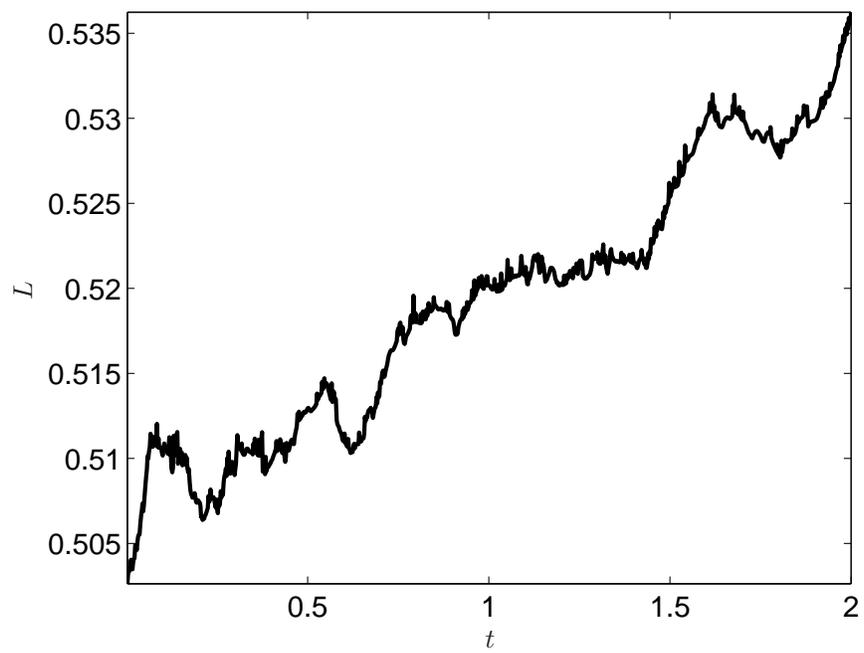}
\end{minipage}
\caption{
Vortex length $L~(\rm cm)$ vs time $t~(\rm s)$ corresponding to 
Fig.~\ref{fig:5}.
}
\label{fig:6}
\end{figure}

We start our analysis of the Kelvin-waves cascade by
noticing that it induces a stretching 
of the vortices:
Fig.~\ref{fig:6} shows that the relative increase of vortex length
is $\Delta L/L \approx 6 \%$. This effect (the transformation
of part of the interaction energy into length) is in agreement with
previous calculations of reconnecting vortex bundles
performed using the Gross-Pitaevskii model
\cite{Alamri,Kerr}. 

\begin{figure}
\begin{minipage}{12cm}
\begin{center}
\includegraphics[width=0.95\textwidth]{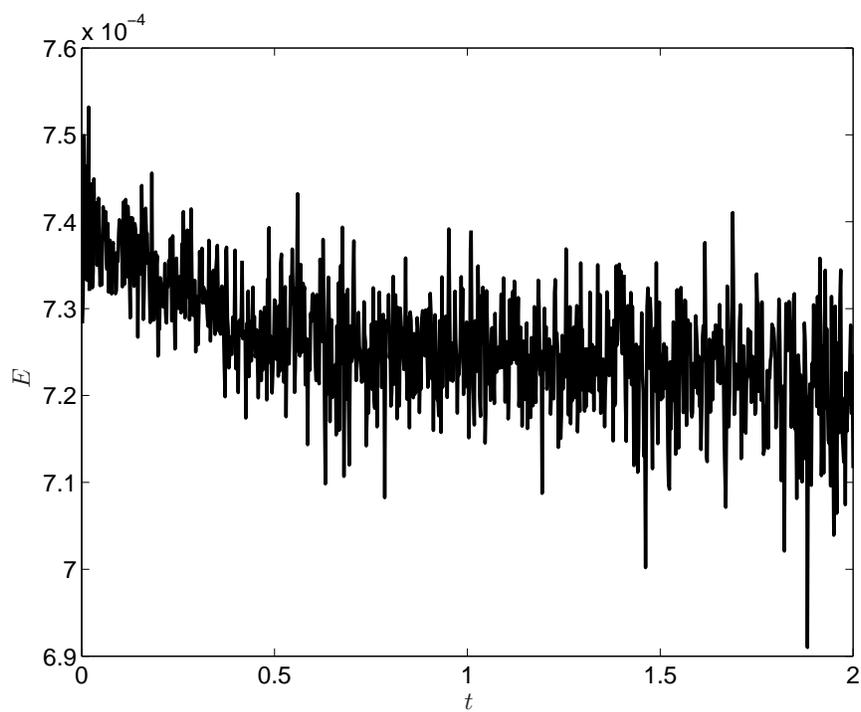}
\end{center}
\end{minipage}
\caption{
Energy $E~({\rm cm^2/s^2})$ vs time $t~({\rm s})$ corresponding to 
Fig.~\ref{fig:5}.
}
\label{fig:7}
\end{figure}

\begin{figure}
\begin{minipage}{12cm}
\begin{center}
\includegraphics[width=0.95\textwidth]{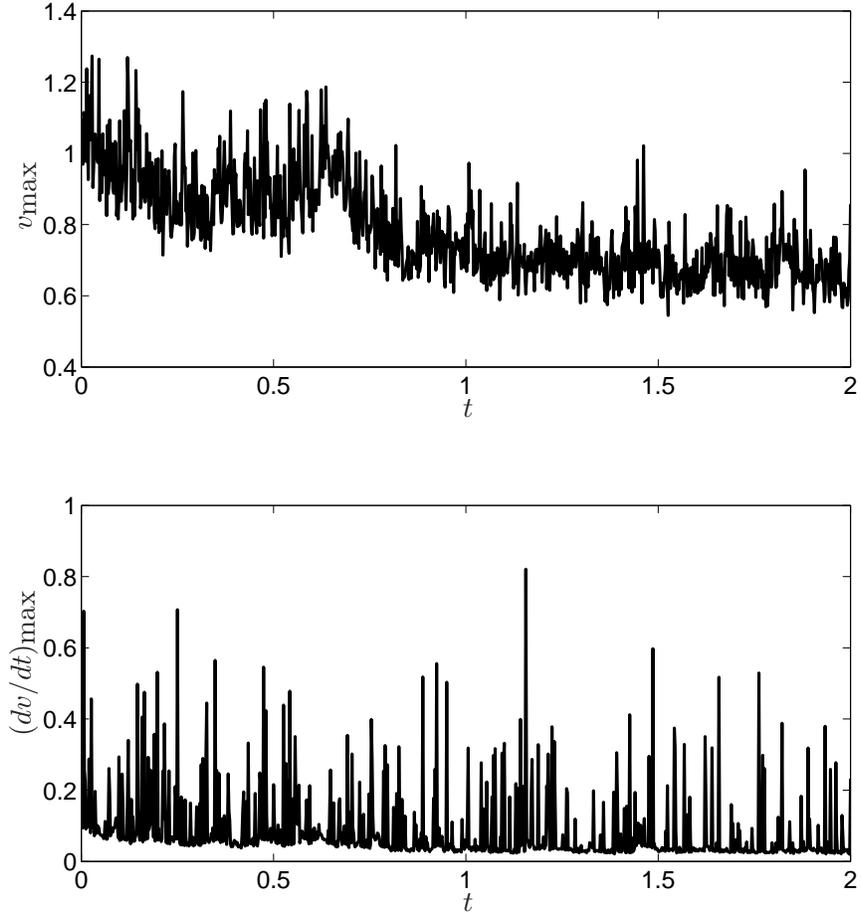}
\end{center}
\end{minipage}
\caption{
Top:
maximum speed $v_{max}~({\rm cm/s})$ vs time $t~({\rm s})$ (top) and
maximum acceleration $(dv/dt)_{max}~({\rm cm/s^2}$) vs $t~({\rm s})$
(bottom) of the vortex points ${\bf s}_j$ $(j=1,\cdots N)$
corresponding to Fig.~\ref{fig:5}.
}
\label{fig:8}
\end{figure}

The behaviour of the energy $E$
is shown in Fig.~\ref{fig:7}. During the initial stage of the
evolution ($0 <t< 0.5~\rm s$) the
energy slightly declines, then it becomes approximately constant,
with relative fluctuations $\Delta E/E$ of less than $3 \%$. 
The energy
fluctuations are larger than for non-cascading vortex configurations,
because in the cascading case
the presence of short, rapid Kelvin waves induces rapid
velocity fluctuations. Fig.~\ref{fig:8}(top) and Fig.~\ref{fig:8}(bottom)
show respectively the maximum speed $v_{max}$ and the maximum 
acceleration $(dv/dt)_{max}$ sampled over the discretization
points ${\bf s}_j$ $(j=1,\cdots N)$. 
It is apparent that velocity fluctuations are larger during
the initial stage of the cascade ($0 < t < 0.5~\rm s$). We shall return
to why the energy initially decreases after discussing the
vortex curvature.

\begin{figure}
\begin{minipage}{12cm}
\includegraphics[width=0.95\textwidth]{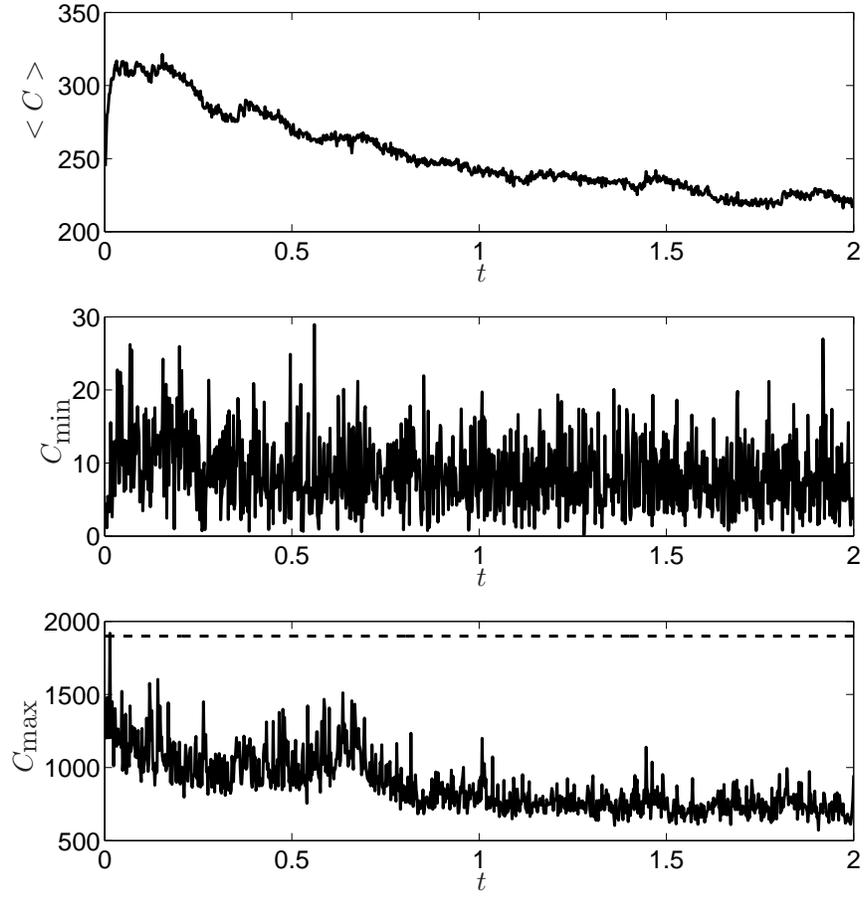}
\end{minipage}
\caption{
Curvature vs time $t$ corresponding to Fig.~\ref{fig:5}.
Top: mean curvature $<C>~\rm (cm^{-1})$ vs $t~\rm (s)$.
Middle: minimum curvature $C_{min}~\rm (cm^{-1})$ vs $t~\rm (s)$.
Bottom: maximum curvature $C_{max}~\rm (cm^{-1})$ vs $t~\rm (s)$; 
the dashed line
is the maximum curvature which corresponds to the numerical
discretization. The maximum curvature which is observed
is therefore always much less than
the numerical resolution.
}
\label{fig:9}
\end{figure}

\begin{figure}
\begin{center}
\includegraphics[width=0.65\textwidth]{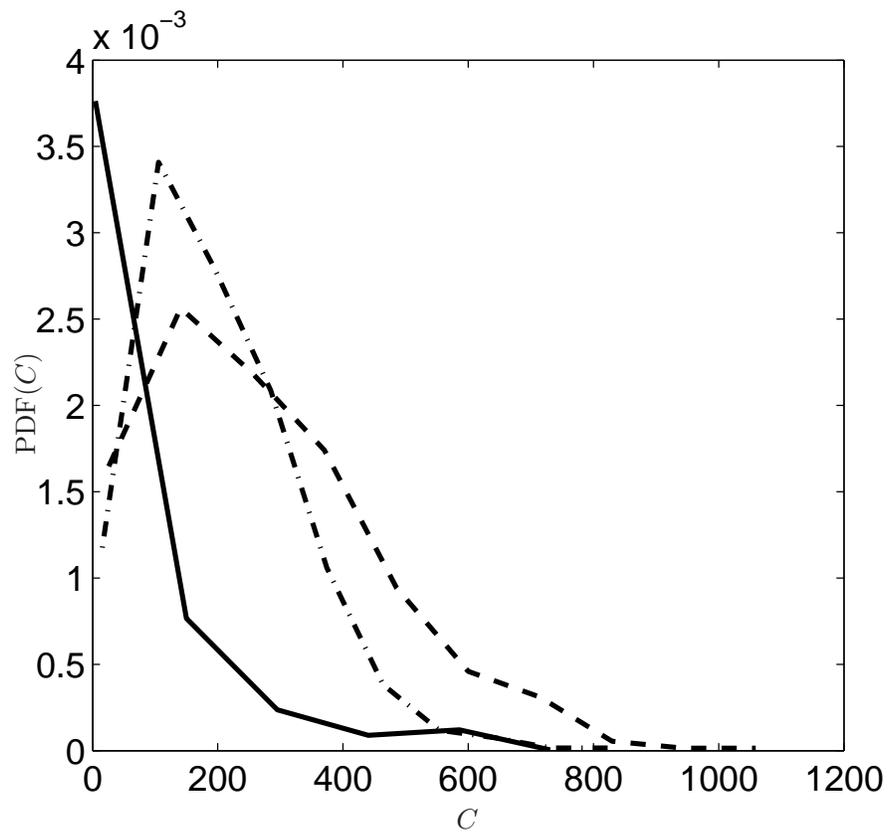}
\caption{
Probability density function of the curvature $\rm PDF(C)$ 
vs $C~(\rm cm^{-1})$ 
corresponding to Fig.~\ref{fig:5} at selected times.
Solid line: at $t=0$.
Dashed line: at $t=0.5~\rm s$.
Dot-dashed line: at $t=1~\rm s$. Note the curve initially moves greatly
to the right (compared to the non-cascading vortices shown in
Fig.~\ref{fig:4}(top))
and then, at later times, slightly to the left (hence lower curvature) as the 
turbulence slowly decays.  
}
\label{fig:10}
\end{center}
\end{figure}

Fig.~\ref{fig:9} shows the time evolution of mean, maximum 
and minimum curvature, $<C>$, $C_{max}$ and $C_{min}$, sampled over
the discretization points $\bs_j$. Fig.~\ref{fig:9}(top) shows that
$<C>$ undergoes a rapid initial rise, then slowly decays, almost
settling down.
Fig.~\ref{fig:9}(bottom) shows that
the maximum curvature is always much less than the maximum value
of curvature allowed by the numerical resolution, which is
$1900~\textrm{cm}^{-1}$ and is represented by the dashed line.
The curvature  PDF 
is shown in Fig.~\ref{fig:10}. Note that, as time progresses, 
${\rm PDF}(C)$
moves to the right towards larger values of $C$. This effect
must be compared against the much smaller
movement to the right of the curvature PDF 
for the non-cascading vortex configuration which is showed
in Fig.~\ref{fig:4}(top). Numerical experiments suggest that
${\rm PDF}(C)$ is indeed a convenient quantity to monitor if one looks
for evidence of the Kelvin-waves cascade.
At later stages, as energy is lost at small scales,
we observe that the turbulence decays, as in the experiments
\cite{McClintock,Bradley},
and the curvature PDF moves to smaller values of $C$.

\begin{figure}
\begin{center}
\includegraphics[width=0.65\textwidth]{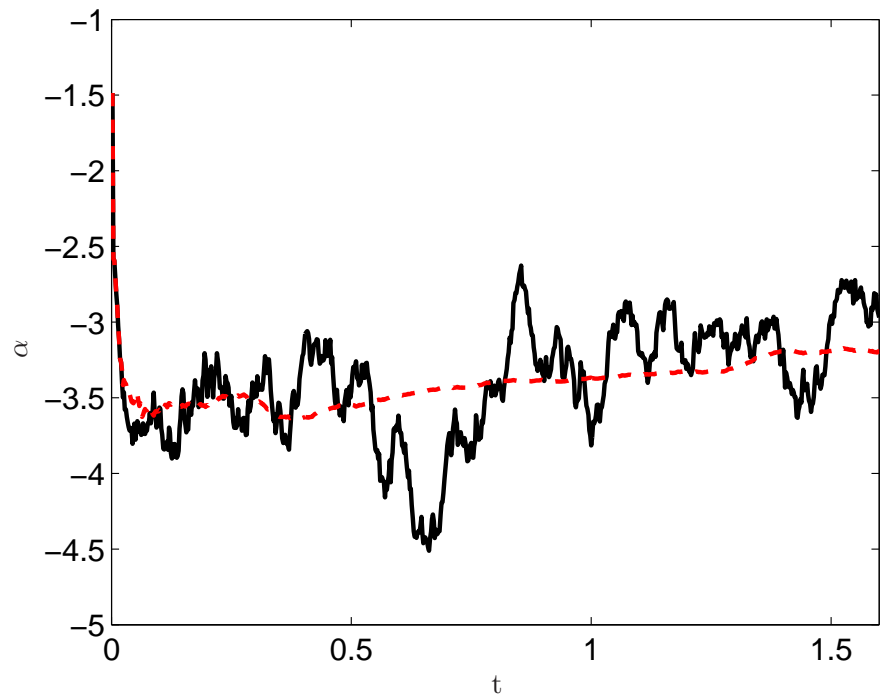}
\caption{(Color online).
Exponent $\alpha$ vs time $t$ of the probability density function
of the curvature, ${\rm PDF(C)} \sim C^{\alpha}$
for large $C$,
corresponding to Fig.~\ref{fig:5}. The (red) dashed line is the
running average.
}
\label{fig:11}
\end{center}
\end{figure}

It is also interesting to note in Fig.~\ref{fig:10} that,
after an initial transient, the curvature
PDF assumes a power law dependence 
${\rm PDF}(C) \sim C^{\alpha}$
for large values of $C$. 
Fig.~\ref{fig:11} shows that the exponent $\alpha$ quickly settles 
down to the average value $\alpha \approx -3.5$.  In the low resolution
run ($\delta=10^{-3}~\rm cm$) a similar calculation gives
$\alpha=-3.34$, which confirms that our results are independent of
the numerical resolution.

\begin{figure}

\begin{minipage}{12cm}
\begin{center}
\includegraphics[width=0.85\textwidth]{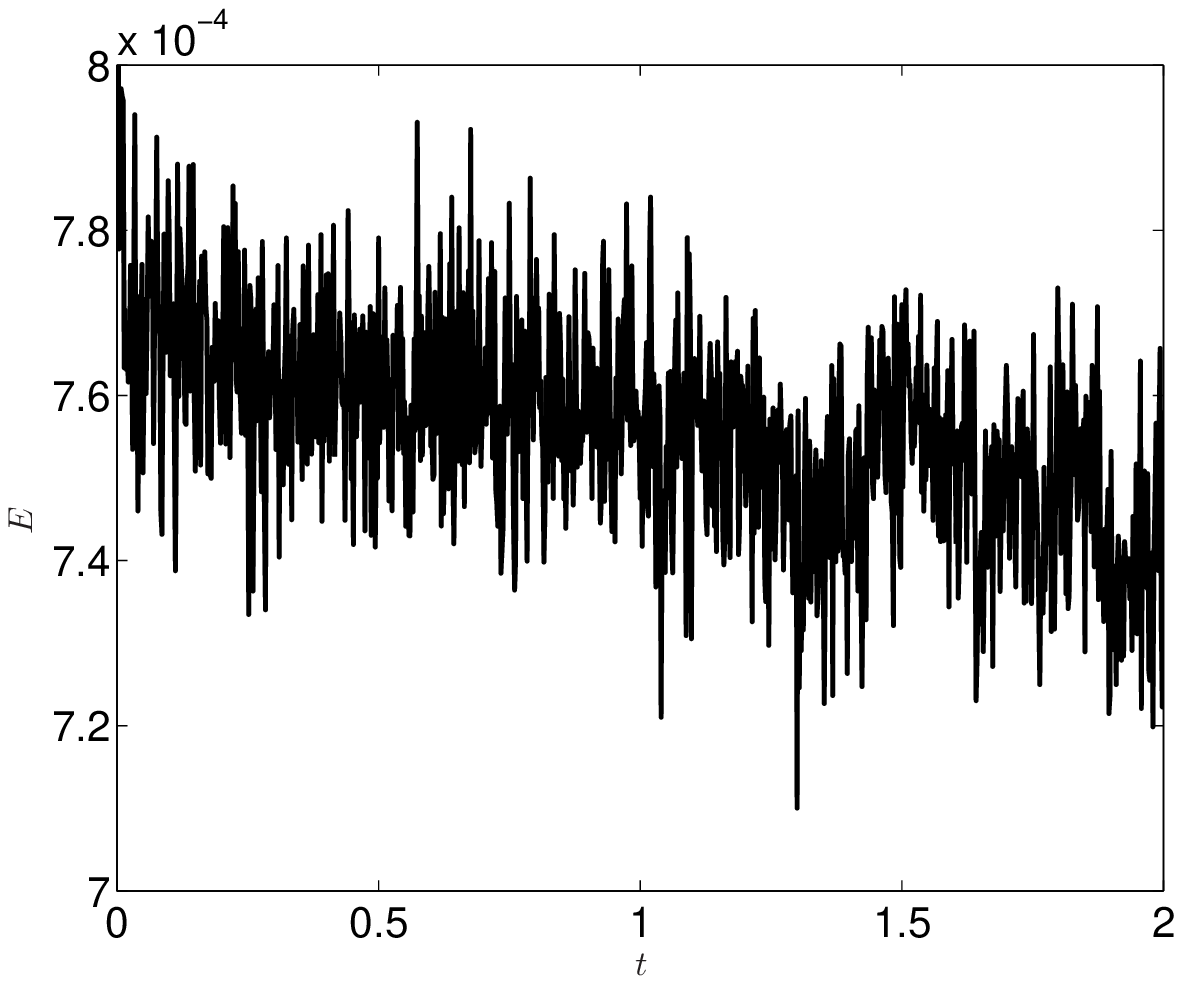}
\end{center}
\end{minipage}

\begin{minipage}{12cm}
\begin{center}
\includegraphics[width=0.95\textwidth]{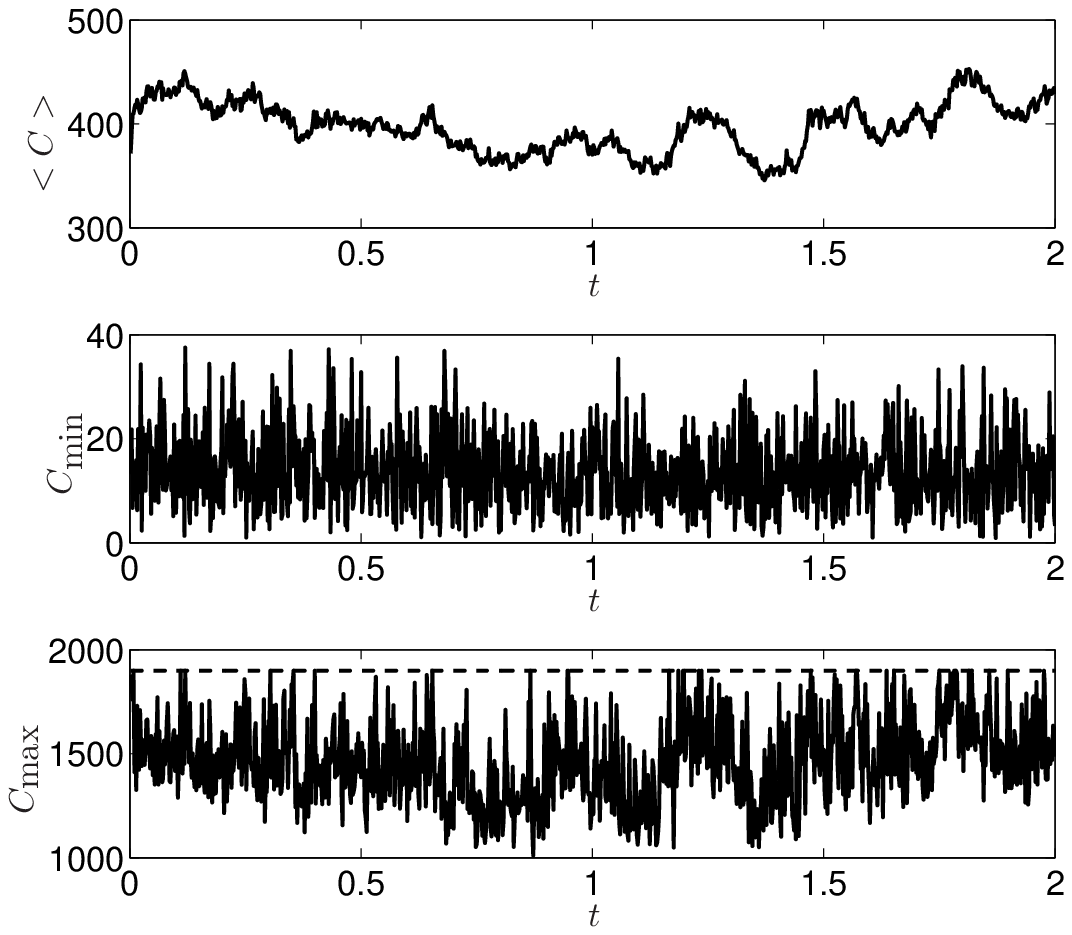}
\end{center}
\end{minipage}

\caption{
Analysis of the evolution of the initial vortex configuration shown in
Fig.~\ref{fig:5} (top left) in the absence of vortex reconnections.
Top:
Energy $E~({\rm cm^2/s^2})$ vs time $t~({\rm s})$.
Bottom: Corresponding mean curvature $<C>$ vs $t$, miniumum curvature $C_{min}$
vs $t$, and maximum curvature $C_{max}$ vs $t$. 
As in Fig.~\ref{fig:9}(bottom), the dashed line is the maximum curvature which 
corresponds to the numerical
discretization.
}
\label{fig:12}
\end{figure}

We can now return to the question as to why there is a small energy loss
during the initial stage of the evolution
from $t=0$ to $t\approx 0.5~\rm s$ in Fig.~\ref{fig:7}.
Our numerical algorithm has two mechanisms
through which length, hence energy, can be lost: vortex reconnections
and smoothing of regions of high curvature. Which is more important ?
To find the answer
we repeat the calculation with the vortex reconnection algorithm switched
off.
The results are shown in Fig.~\ref{fig:12}. If we compare the energy
computed with reconnections (Fig.~\ref{fig:7}) and 
without reconnections (Fig.~\ref{fig:12}(top))
we see that $E$ is more constant without reconnections. It is also
instructive to compare
curvature statistics computed with reconnections
(Fig.~\ref{fig:9})
and without reconnections (Fig.~\ref{fig:12}, bottom three graphs). 
It is apparent that
vortex reconnections are the most important energy
sink at large $k$: without reconnections, the maximum curvature
$C_{max}$ is much larger, and often reaches the numerical limit represented
by the horizontal dashed line in Fig.~\ref{fig:12}(bottom). 
Since $C_{min}$ is approximately the same, $<C>$ is also
larger than in the reconnecting case.
Therefore, without vortex reconnections the cascade proceeds to the smallest 
scale which is numerically allowed, at which point the smoothing algorithm 
kicks in.
We conclude that the initial energy loss shown in Fig.~\ref{fig:7} 
is  due to the
frequent reconnections which occur during the initial stage of the
evolution.

\subsection{Spectrum}\label{III.C}

\begin{figure}
\begin{center}
\includegraphics[width=0.65\textwidth]{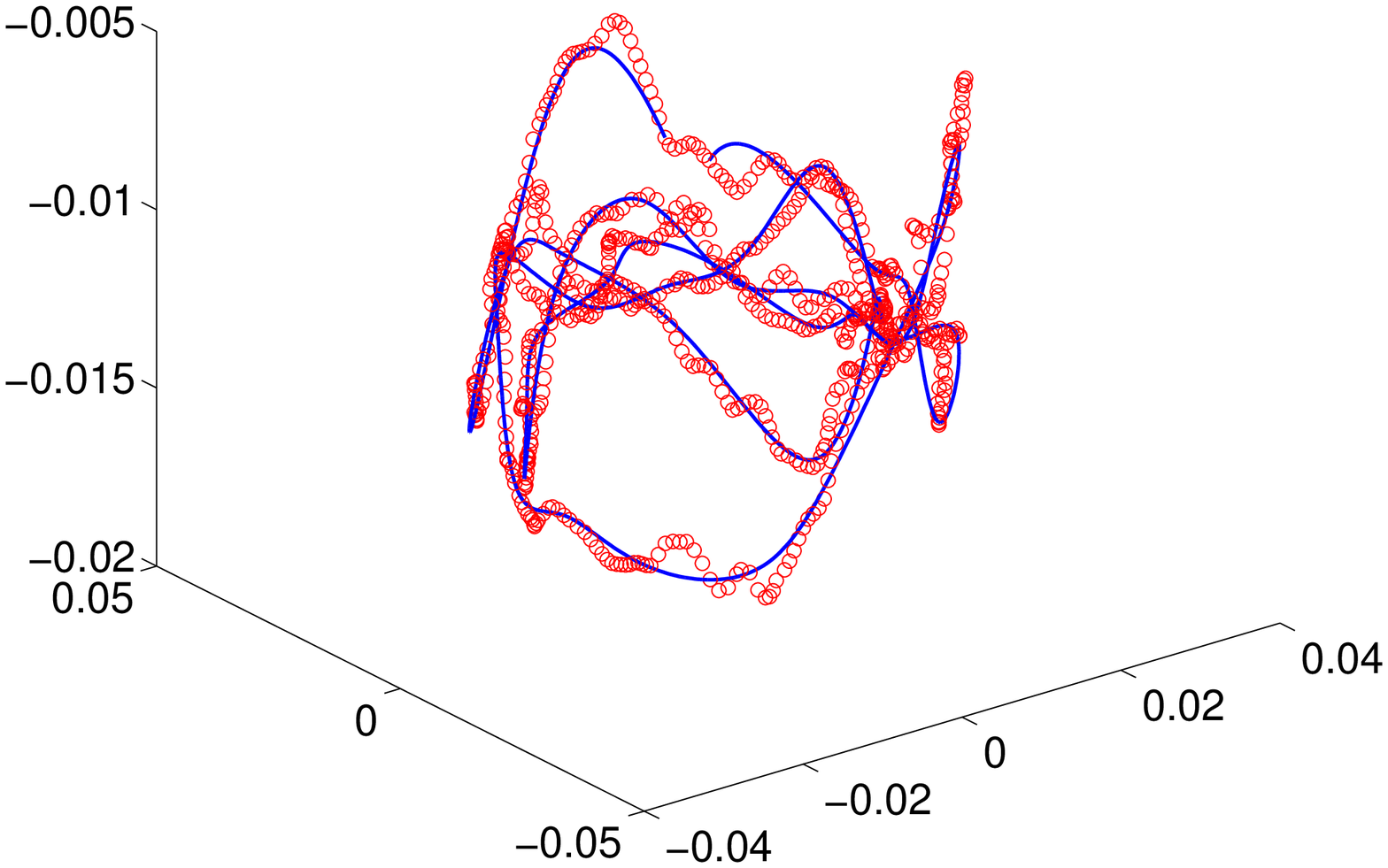}
\includegraphics[width=0.65\textwidth]{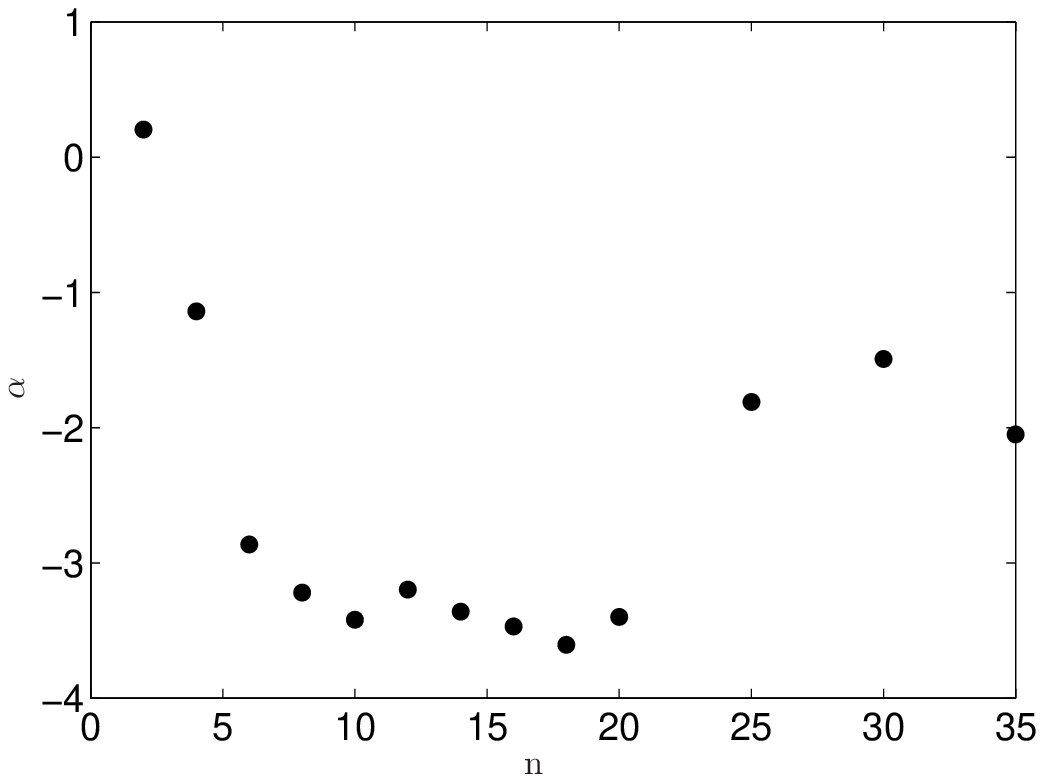}
\caption{
(Color online).
Top: Smoothed vortex configuration. 
The (red) circles denote the actual discretization
points $\bs_j$ ($j=1,\cdots N$) of the vortex configuration
corresponding to Fig.~\ref{fig:5} at time $t=0.18~\rm s$. 
The (blue) line is the smoothed curve
curve $\bs_{smooth}$ obtained by cubic spline fit interpolation
over every $n^{\rm th}=15$ points.
Bottom: Slope $\beta$ of the amplitude spectrum $A(k) \sim k^{\beta}$
for large $k$ plotted as a function of $n$.
Note the plateau $10 < n < 20$ where the spectral slope is roughly constant. 
}
\label{fig:13}
\end{center}
\end{figure}

The next quantity of interest is the amplitude
$a(\xi)$ of the Kelvin waves and the spectrum of the Kelvin-waves
cascade. 
Following Svistunov \cite{Svistunov}, we introduce the
concept of {\it smoothed vortex filament}: at every time $t$,
we use every $n=15$ points as nodes of a cubic--spline 
interpolation to obtain a new (smoothed) filament,  $\bs_{\rm smooth}$
from the original discretised vortex filament $\bs_j$ $(j=1,\cdots N)$. 
Fig.~\ref{fig:13} (top) is an example of this procedure: in the
figure we see the
actual discretization points ${\bf s}_j$ $(j=1,\cdots N)$
together with the smoothed line. We choose
$n=15$ because, in a range which is approximately
$10 < n<20$, the amplitude spectrum (which we shall define shortly)
does not depend on $n$, as shown in Fig.~\ref{fig:13}(bottom),
confirming that our numerical smoothing
procedure is robust enough.

\begin{figure}
\begin{center}
   \includegraphics[width=0.65\textwidth]{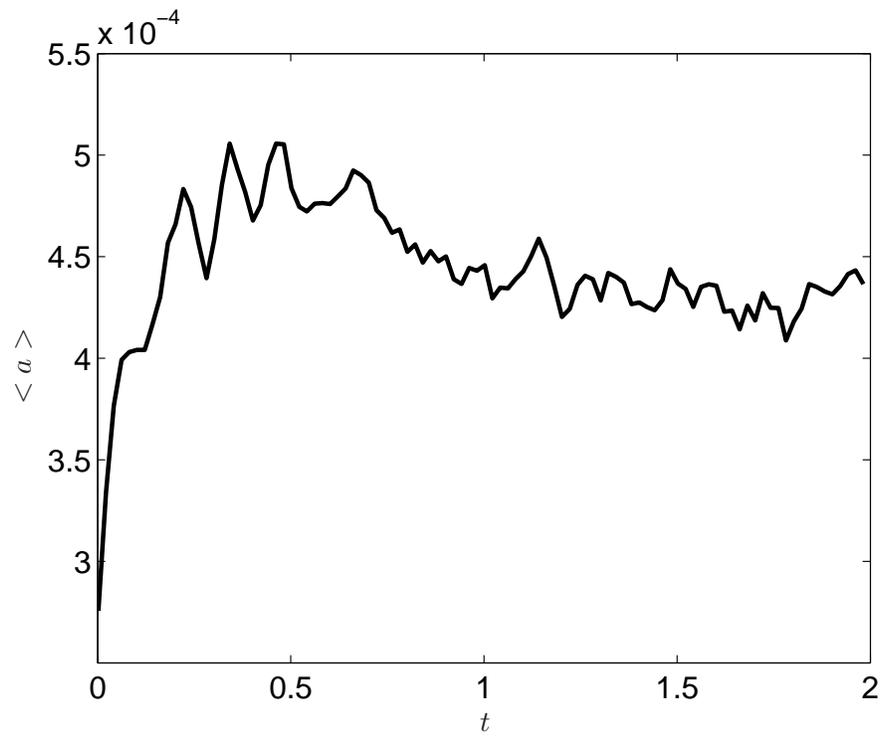}
   \caption{
Mean amplitude $<a>~(\rm cm)$ of the Kelvin waves vs time $t~{\rm s}$
corresponding to Fig.~\ref{fig:13}. Note the saturation for $t>1~(\rm s)$.
}
\label{fig:14}
\end{center}
\end{figure}

We then define the Kelvin wave amplitude $a(\xi)$
as the distance between the original filament  and the smoothed
filament: $a(\xi)=\vert \bs-\bs_{\rm smooth} \vert$. The 
mean amplitude $<a>$ as a function of time is shown in Fig.~\ref{fig:14}.
Note that the mean amplitude $<a>$ saturates for $t>1~{\rm s}$.

\begin{figure}
\begin{center}
   \includegraphics[width=0.65\textwidth]{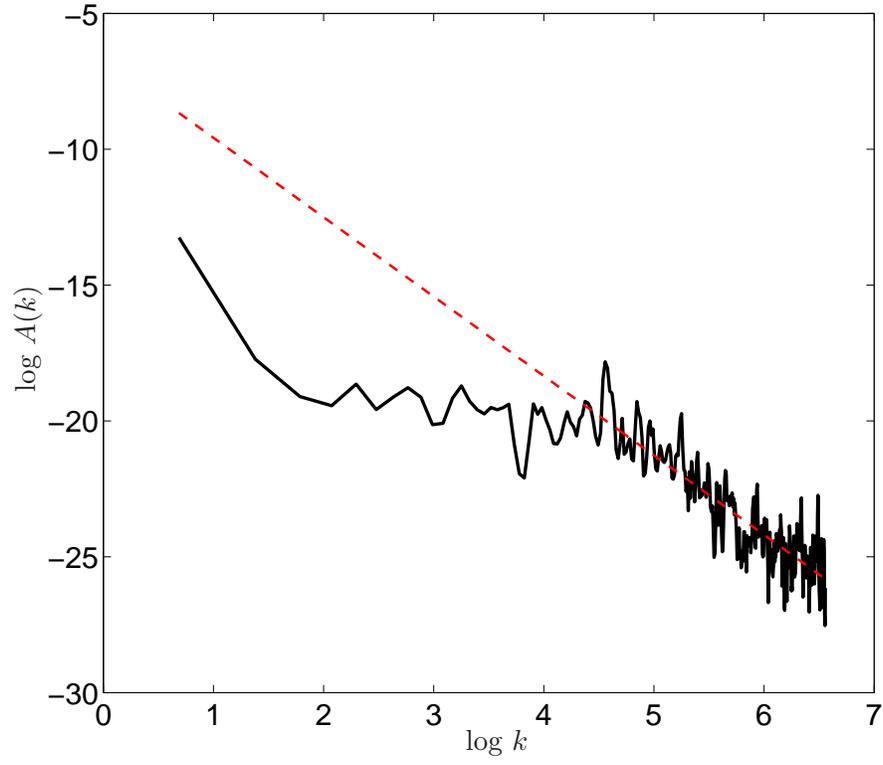}
   \caption{
(Color online)
Log log plot of Kelvin wave amplitude spectrum
$A(k)$ ($\rm cm^4$) vs wavenumber $k$ ($\rm cm^{-1}$) at time
$t=0.18~(\rm s)$, corresponding the  Fig.~\ref{fig:13}.
The (red)
dashed line represents the large-k behaviour
$A(k) \sim k^{\beta}$ with the best fit $\beta=-3.1$
}
\label{fig:15}
\end{center}
\end{figure}

\begin{figure}
\begin{center}
\includegraphics[width=0.65\textwidth]{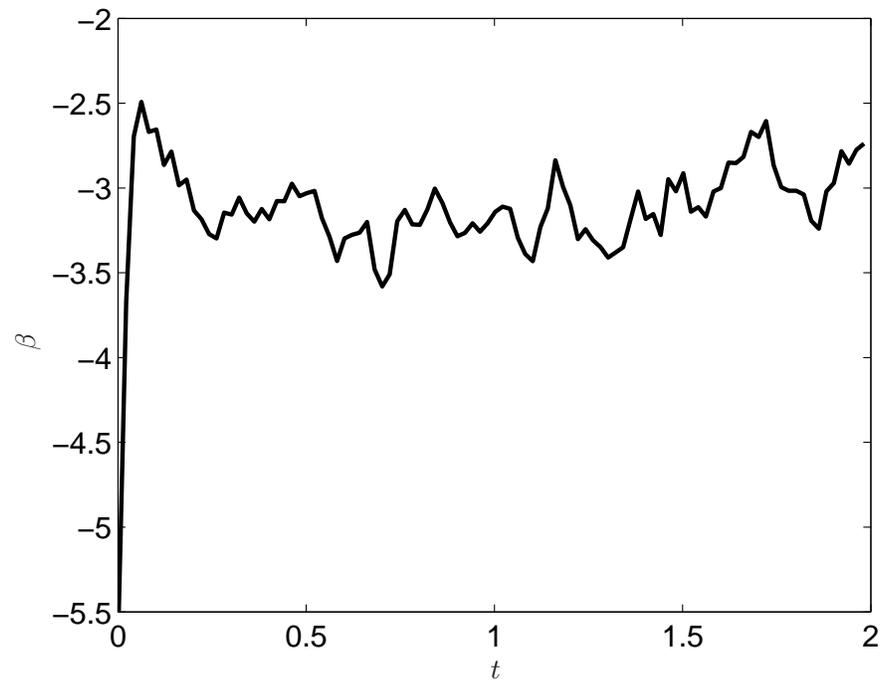}
\caption{
Slope $\beta$ of the large-k amplitude spectrum $A(k) \sim k^{\beta}$
vs time $t~{\rm s}$. Note the
quick saturation.
}
\label{fig:16}
\end{center}
\end{figure}

Finally, the amplitude spectrum, $A(k)$, is defined as

\begin{equation}
\frac{1}{2}\int a^2(\xi) d\xi=\int_0^\infty A(k') dk',
\end{equation}

\noindent
and is shown in Fig.~\ref{fig:15}.
We find that the amplitude spectrum behaves as
$A(k) \sim k^\beta$ for large $k$. 
The best fit is $\beta=-3.10$; we obtain $\beta=-3.21$ in the
low resolution run.
Fig.~\ref{fig:16} shows that the scaling exponent $\beta$ is
independent of time, as the Kelvin-waves cascade develops. 

\subsection{Angle}\label{III.D}

\begin{figure}
\begin{minipage}{12cm}
%\begin{center}
   \includegraphics[width=0.7\textwidth]{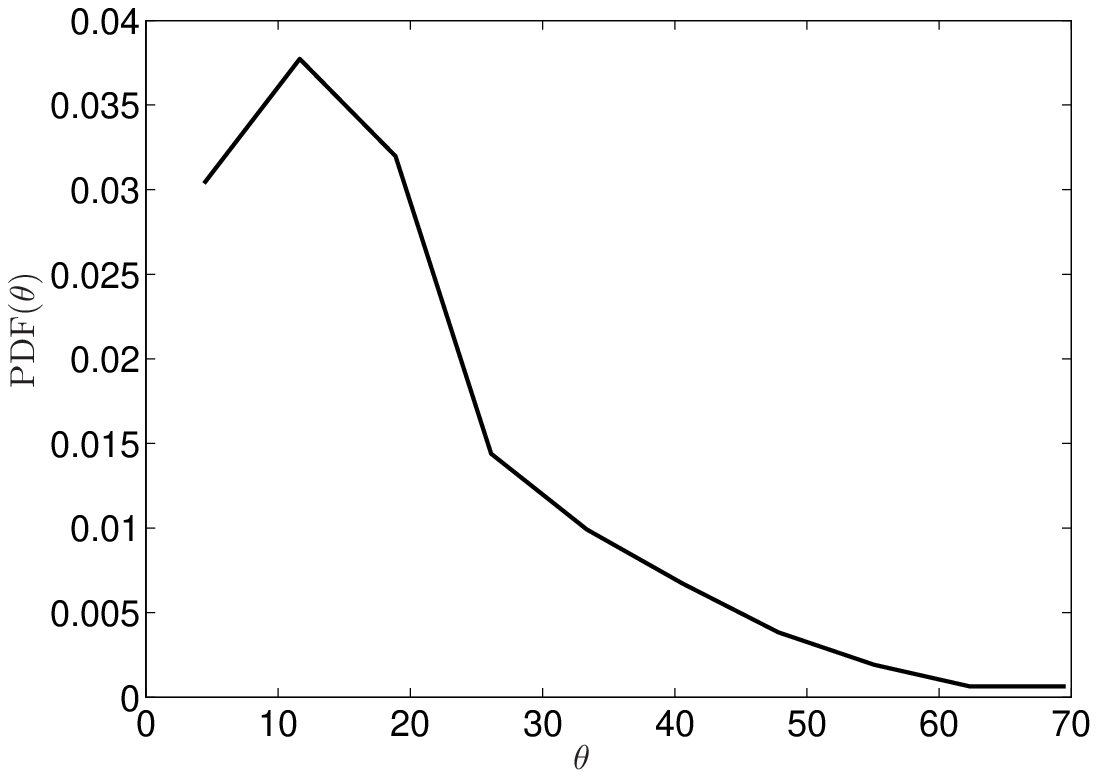}
%\end{center}
\end{minipage}

\begin{minipage}{12cm}
%\begin{center}
   \includegraphics[width=0.75\textwidth]{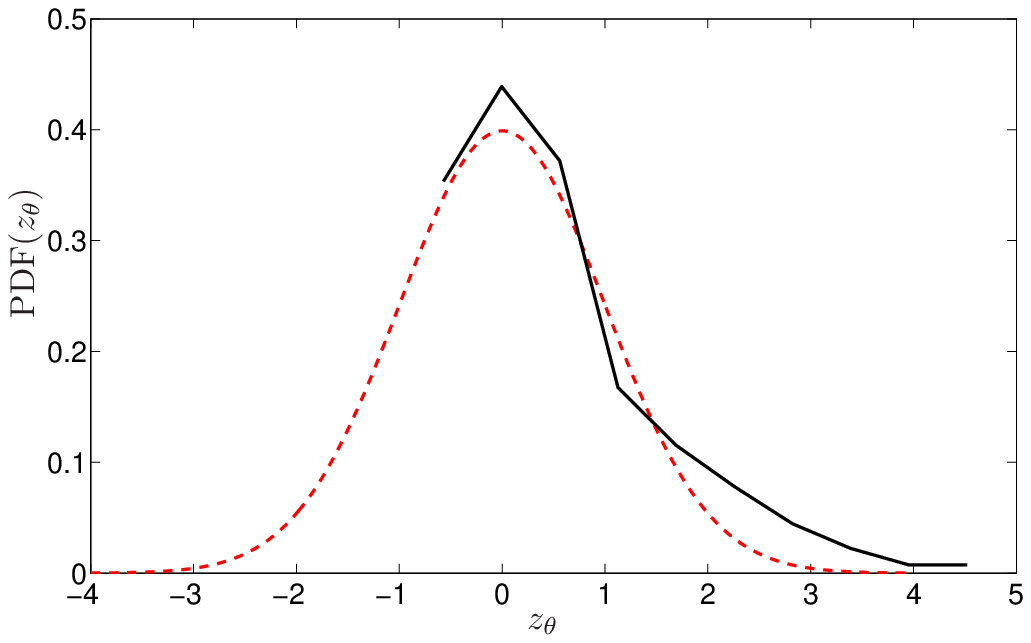}
%\end{center}
\end{minipage}
\caption{
(Color online).
Angle between vortex filament and smoothed vortex filament.
Top:
Probability density function ${\rm PDF}(\theta)$ vs $\theta$
of the angle $\theta$ (in degrees)
between the smoothed vortex filament and the 
actual vortex filament at $t=0.2~{\rm s}$.
Bottom:
Probability density function ${\rm PDF}(z_{\theta})$ vs $z_{\theta}$
of the scaled angle $z_\theta=(\theta-\bar{\theta})/\sigma(\theta)$
where $\bar{\theta}$ and $\sigma(\theta)$ are respectively
the mean and the variance
of the distribution of the angle $\theta$.
The dashed (red) line shows the normal PDF. Note 
the slight departure from Gaussianity at large angles.
}
\label{fig:17}
\end{figure}

From the dot product of the local tangent vectors, 
$\bs'$ and $\bs'_{\rm smooth}$, at each time $t$ we calculate 
the angle $\theta$ between the original vortex filament $\bs$ and
the smoothed filament $\bs_{\rm smooth}$.   
The normalised histogram, ${\rm PDF}(\theta)$, computed at $t=0.002~\rm s$
is shown in Fig.~\ref{fig:17}(top). The peak value
is approximately at $\theta \approx 15^o$. 

In order to compare the distribution which we find with that of a standard 
normal distribution (zero mean and unit standard deviation), we make use of 
the following standard result from statistics.
All normal random variables with mean $\mu$ and variance $\sigma$ can be 
`standardised' to fit a standard normal distribution by subtracting 
the mean and scaling this quantity by the standard deviation.
Fig.~(\ref{fig:17})(bottom)
shows a plot of $z_\theta=(\theta-\bar{\theta})/\sigma(\theta)$
where $\bar{\theta}$ and $\sigma(\theta)$ are respectively
the mean and the variance
of the distribution of the angle $\theta$.
%(dashed and dotted lines
%for $t=0.02~\rm s$ and $t=0.12~\rm s$ respectively) 
It is apparent that the distribution of the angles is close to a normal distribution. This result
validates the weak-turbulence theory approach to the Kelvin-waves
cascade problem\cite{Sheko}. A small departure from Gaussianity is
present at large angles; it is likely that this effect is due to the
fractal nature of the vortex filament (see section~\ref{III.F}).

\subsection{Reconnecting vortex rings}\label{III.E}

\begin{figure}
\begin{minipage}{8cm}
\includegraphics[width=0.95\textwidth]{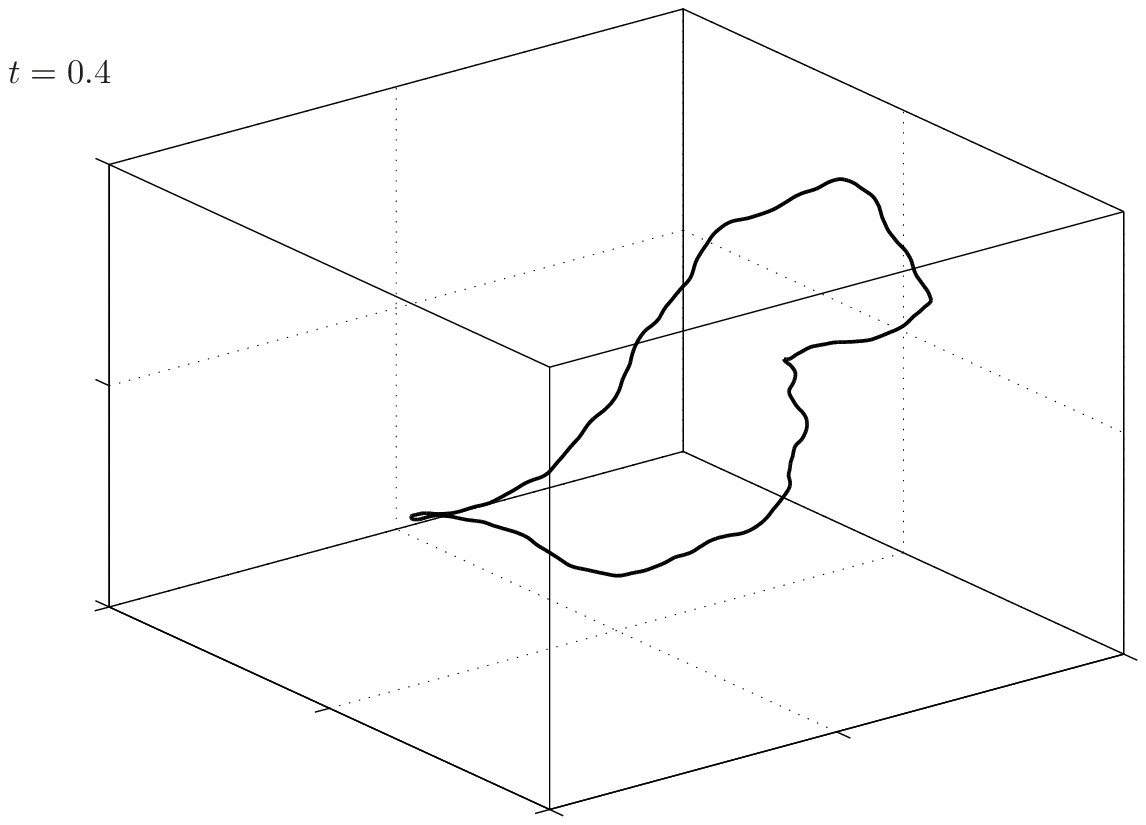}
\end{minipage}
\begin{minipage}{8cm}
\includegraphics[width=0.95\textwidth]{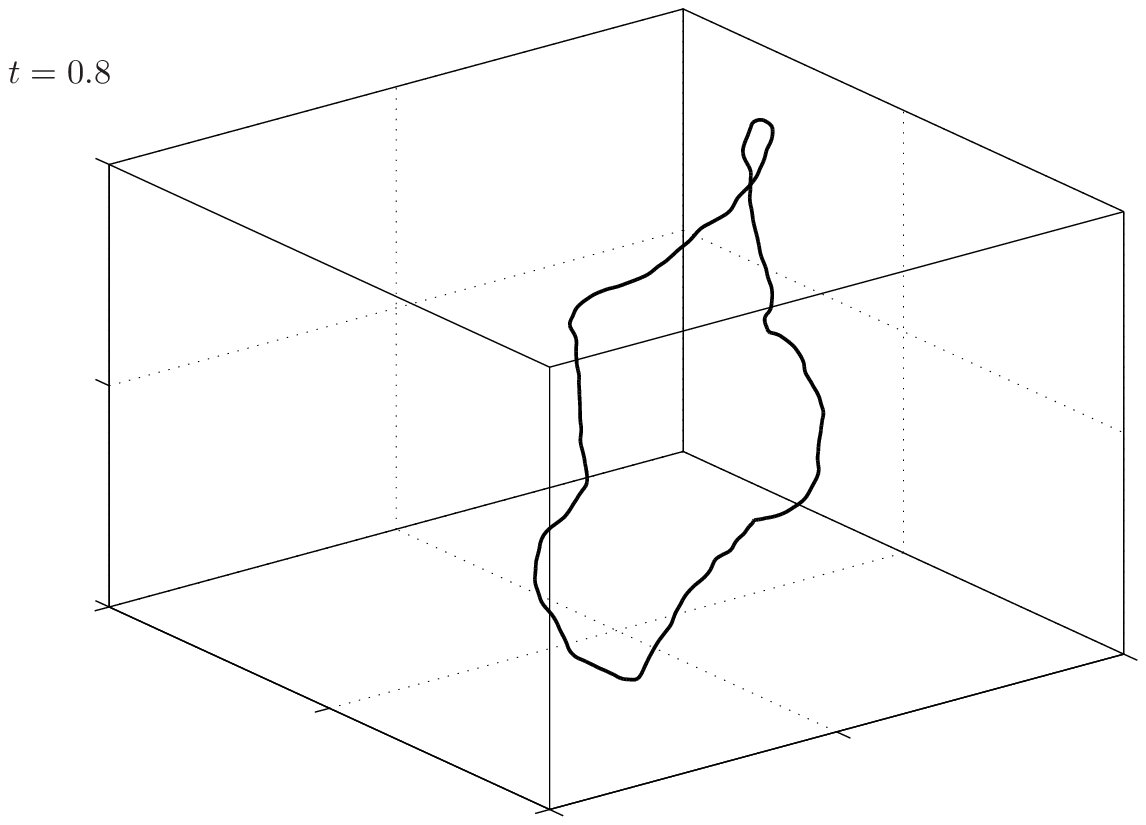}
\end{minipage}
\caption{
Non-cascading reconnecting vortex rings. Time evolution of two rings of
radius $R=0.16 \rm cm$ initially linked to each other. Note that
the reconnection cusp relaxes and induces Kelvin waves, but there
is no cascade.
Top left: At time $t=0~\rm s$;
Top right: $t=0.16~\rm s$;
Bottom left: $t=0.5~\rm s$;
Bottom right: $t=0.8~\rm s$.
}
\label{fig:18}
\end{figure}

\begin{figure}

\begin{minipage}{10cm}
\begin{center}
\includegraphics[width=0.85\textwidth]{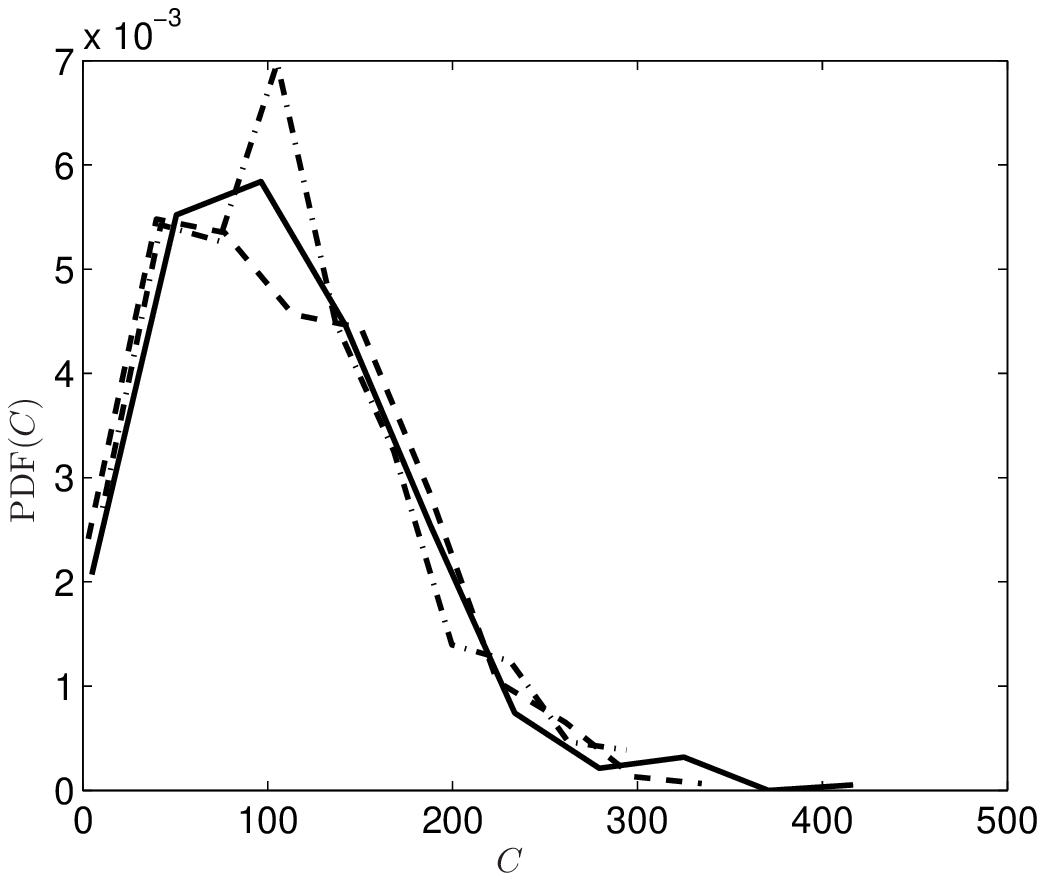}
\end{center}
\end{minipage}

\begin{minipage}{10cm}
\begin{center}
\includegraphics[width=.95\textwidth]{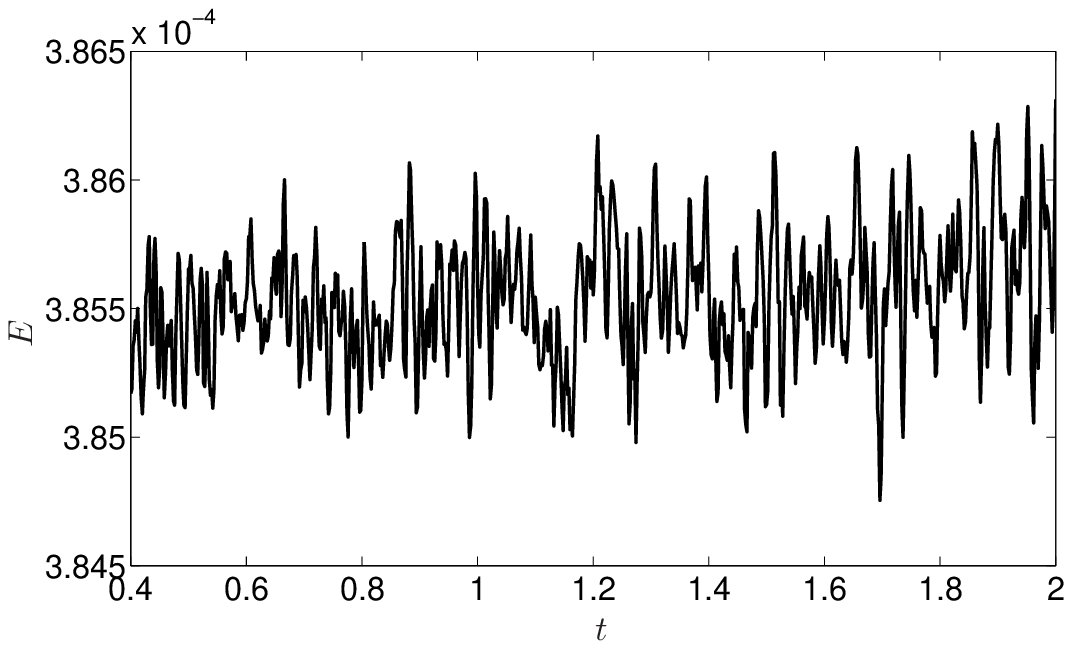}
\end{center}
\end{minipage}

\caption{
Evolution of curvature and energy corresponding to Fig.~\ref{fig:18}.
Top:  
${\rm PDF}(C)$ vs $C~(\rm cm^{-1})$ 
at $t=0$ (solid line), $t=0.2~\rm s$ (dashed line) and
$t=1.6~\rm s$ (dot dashed line) Note that the peak of the PDF does not move to
larger values of $C$.
Bottom: Corresponding behaviour of the energy 
$E~(\rm cm^2/s^2)$ vs time $t~(\rm s)$.
}
\label{fig:19}
\end{figure}

The last set of numerical experiments is concerned with
two vortex rings which reconnect with each other. This configuration is
similar to the collision of four vortex rings which was used in
a previous work\cite{Kivotides-cascade}.
Fig.~\ref{fig:18} shows the time evolution. It is apparent that,
over the time scale which is considered, no Kelvin-waves cascade develops.
This visual conclusion is confirmed by the inspection of 
${\rm PDF}(C)$, shown in Fig.~\ref{fig:19}(top), which 
does not progress in time  to larger values of $C$, unlike Fig.~\ref{fig:10}. 
Fig.~\ref{fig:19}(bottom) shows that the energy remains constant
within $\Delta E/E \sim \pm 0.12 \%$.

\begin{figure}
\begin{minipage}{8cm}
\includegraphics[width=0.95\textwidth]{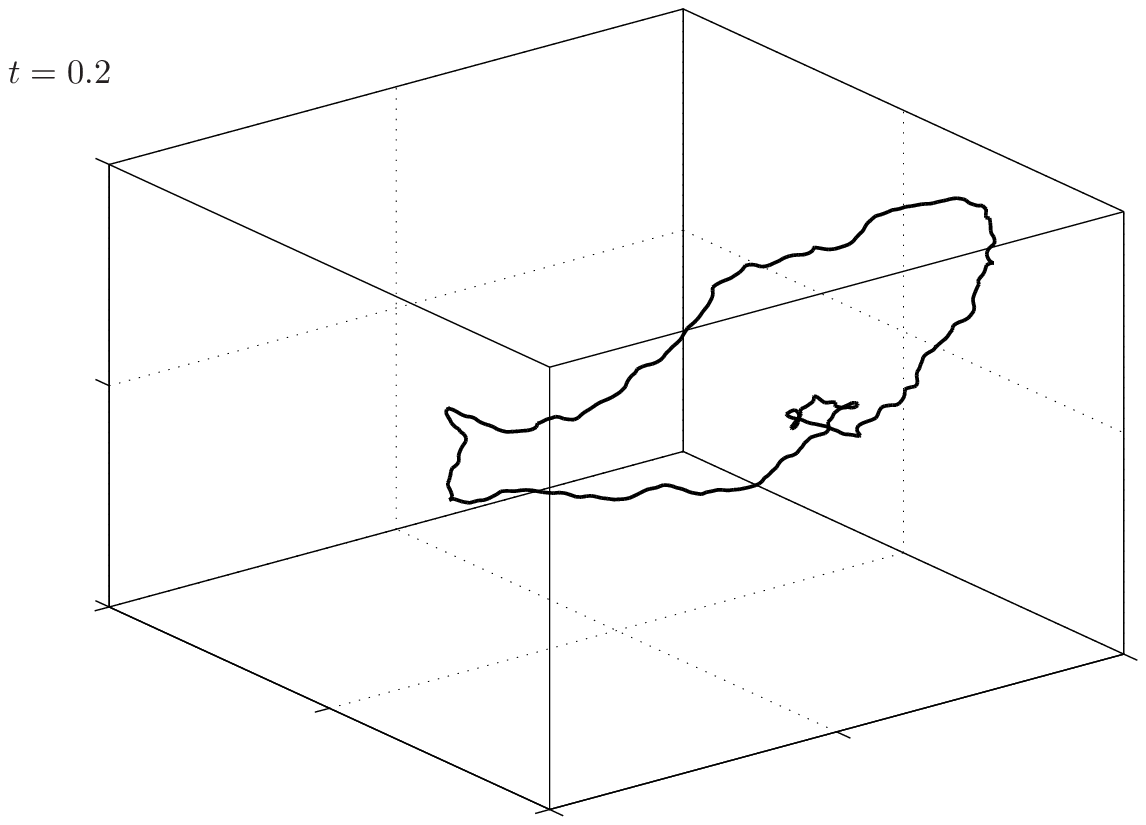}
\end{minipage}
\begin{minipage}{8cm}
\includegraphics[width=0.95\textwidth]{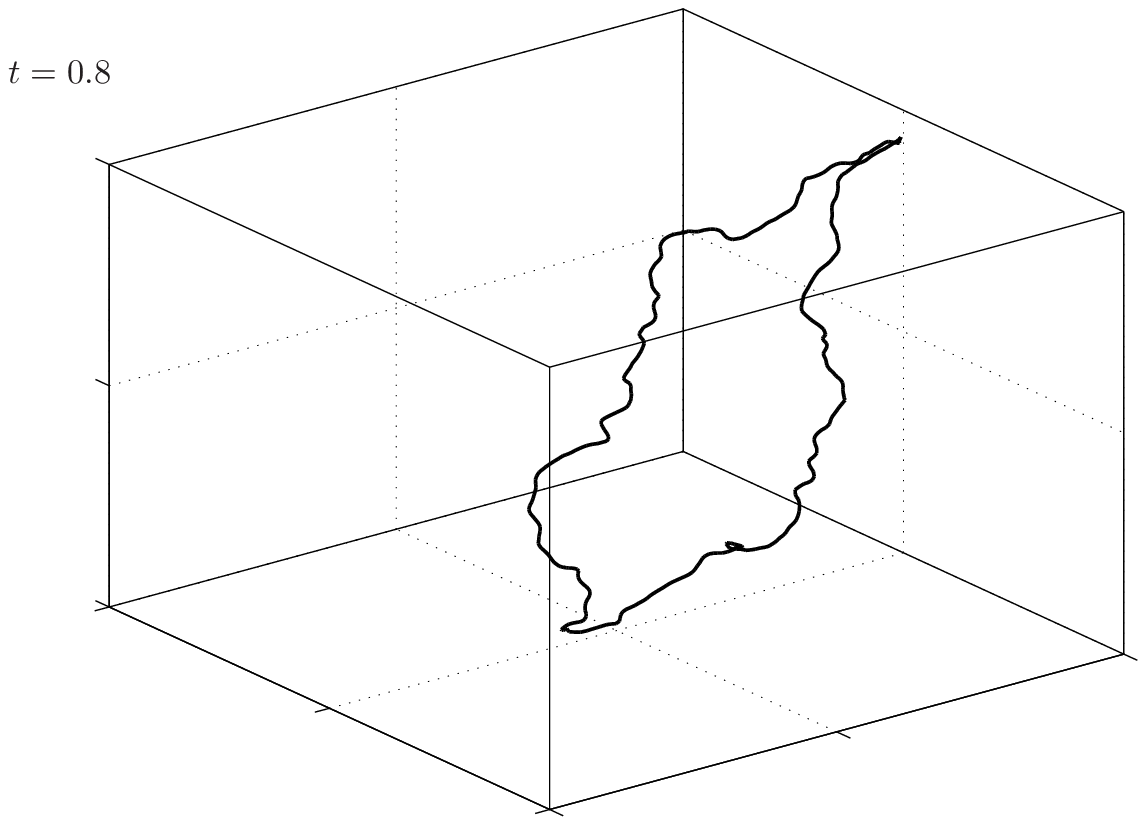}
\end{minipage}
\caption{
Cascading reconnecting vortex rings. Time evolution of two rings of
radius $R=0.16~\rm cm$ initially linked to each other.
This initial configuration differs from the configuration
of Fig.~(\ref{fig:15}) only because some Kelvin waves have been superimposed
to the vortex rings. Note the development of the Kelvin waves cascade.
Top left: At time $t=\rm s$;
Top right: $t=~\rm s$;
Bottom left: $t=~\rm s$;
Bottom right: $t=~\rm s$.
}
\label{fig:20}
\end{figure}

If the initial vortex rings are not perfectly circular, however, the
evolution proceeds toward a Kelvin-waves cascade, as shown in
Fig.~\ref{fig:20}. The increasing wiggliness of Fig.~\ref{fig:20}
must be compared against Fig.~\ref{fig:18}.
Following the initial reconnection of the two rings, we observe $4$ more
reconnections, which are actually self-reconnections since there is only
one vortex filament left. Fig.~\ref{fig:20}(c) shows   
a self-reconnection event which arises from a large twisting (supercoiling)
of the vortex filament, and which results
in the emission of a small vortex loop.
This self-crossing scenario was predicted by 
Kozik and Svistunov\cite{Kozik-Svistunov-PRB-2008}. 
Similar small-loop emission were
also observed in numerical studies of vortex bundle dynamics \cite{Alamri}.

\begin{figure}

\begin{minipage}{10cm}
\begin{center}
\includegraphics[width=0.9\textwidth]{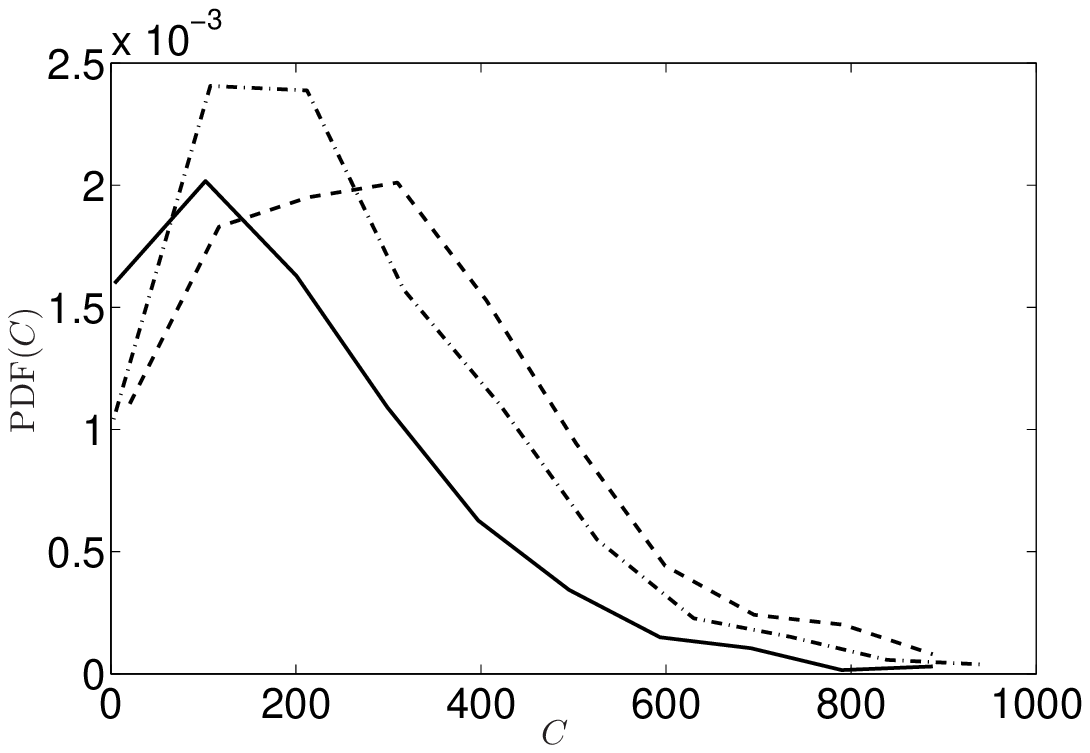}
\end{center}
\end{minipage}

\begin{minipage}{10cm}
\begin{center}
\includegraphics[width=0.95\textwidth]{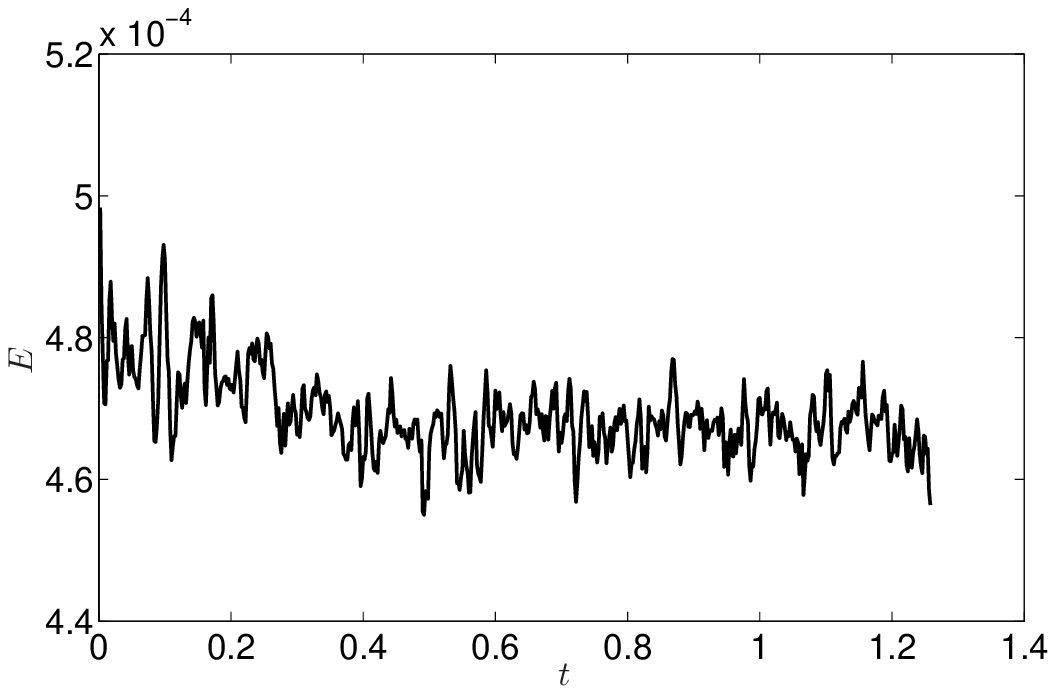}
\end{center}
\end{minipage}

\caption{
Top: Probability density function of the curvature 
${\rm PDF}(C)$ vs $C~(\rm cm^{-1})$ corresponding to Fig.~(\ref{fig:20})
at $t=0$ (solid line), $t=0.2~\rm s$ (dashed line) and
$t=0.8~\rm s$ (dot dashed line) Note that the peak moves to
larger values as in Fig.~\ref{fig:10}.
Bottom: Corresponding behaviour of the energy 
$E~(\rm cm^2/s^2)$ vs time $t~(\rm s)$.
}
\label{fig:21}
\end{figure}

The PDF of the curvature confirms the existence of a cascade: 
Fig.~\ref{fig:21}(top) shows that ${\rm PDF}(C)$
progresses in time to larger values of $C$. 
Fig.~\ref{fig:21}(bottom) shows the behaviour of the energy, which has
the same slight initial decreases as in Fig.~\ref{fig:7} which we have
discussed.

\subsection{Fractal dimension}\label{III.F}

\begin{figure}
\begin{center}
\includegraphics[width=0.75\textwidth]{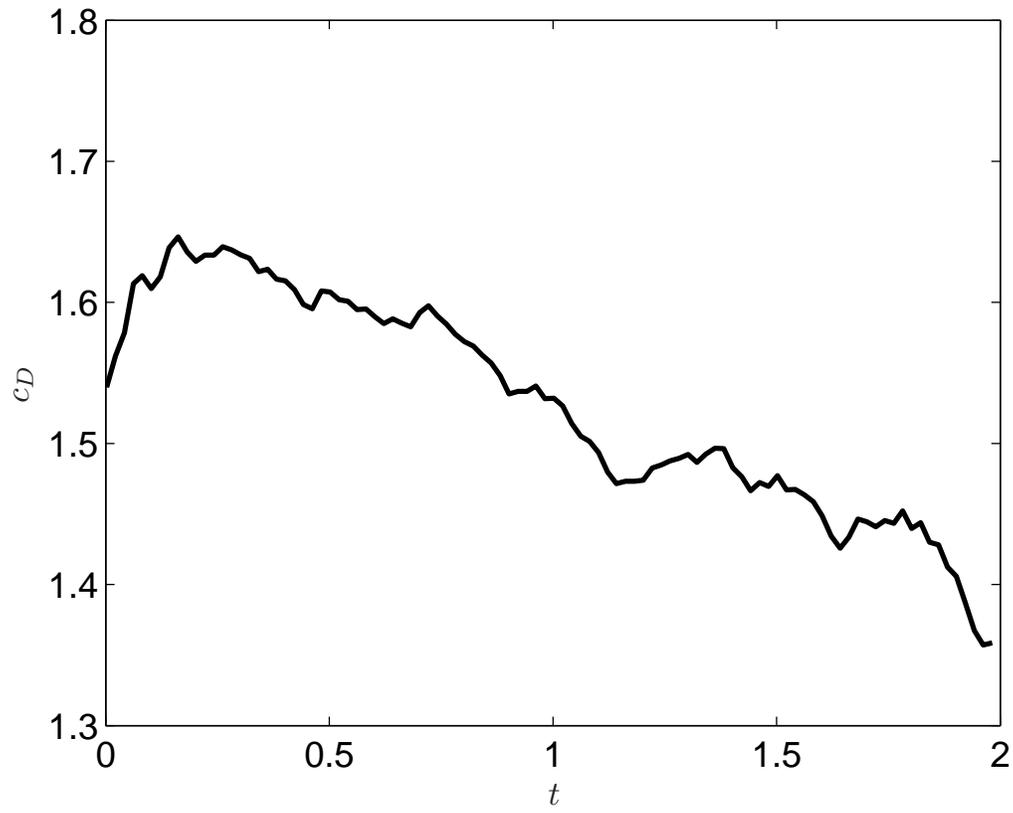}
\end{center}
\caption{
Correlation dimension $c_D$ as a function of time $t~\rm (s)$
corresponding to Fig.~\ref{fig:20}. 
Note a rapid increase of $c_D$ after the 
reconnection at $t=0.008 \rm s$, and the slow decrease which follows
due to numerical dissipation.
}
\label{fig:22}
\end{figure}

To quantify the wiggliness of the vortex configuration
undergoing the Kelvin-waves
cascade,  we compute the correlation dimension $c_D$ introduced by
Grassberger and Procaccia \cite{Grassberger}. The correlation
dimension provides
an upper--bound estimate of the fractal dimension 
The definition is the following:
given the $N$ discretization points $\bs_j$ $(j=1,\cdots N)$,
we define the correlation integral as $K(\epsilon)=n/N^2$ 
where $n$ is the number of points whose separation is less than $\epsilon$.
In the limit of $\epsilon \to 0$, $K(\epsilon)$ takes the form,
\begin{equation}
K(\epsilon)\sim \epsilon^{c_D},
\end{equation}

\noindent
where $c_D$ is the correlation dimension.
This quantity can be calculated efficiently at the same time as the
Biot-Savart integral is calculated. The result is shown in Fig.~\ref{fig:22}
for reconnecting vortex rings undergoing the Kelvin-waves cascade.
Note the rapid increase of $c_D$ during the initial stage caused by 
reconnection events, followed by a decline due to the numerical dissipation
at high wavenumbers which models physical dissipation (sound emission).

\section{Discussion}\label{IV}

Our numerical calculations show that the Kelvin-waves cascade can be
initiated by the interaction of neighbouring vortex lines as well
as by vortex reconnections.
We find that, shortly after an initial transient,
the Kelvin-waves cascade quickly settles in and 
saturates: Fig.~(\ref{fig:11}) and Fig.~(\ref{fig:16})
show that the slopes $\alpha$ and $\beta$
of the curvature histogram at large $C$,
${\rm PDF}(C)\sim C^{\alpha}$, and of the amplitude spectrum at large $k$, 
$A(k)\sim k^{\beta}$, remain approximately constant.

We also find that the Kelvin-waves amplitude spectrum scales approximately as
$A(k) \sim k^{-3.1}$ for large $k$, in agreement
with $A \sim k^{-3}$ reported in Ref.~\cite{Vinen-cascade} for a continuously
excited vortex line, and less steep but in fair agreement with the
theories of L'vov, Nazarenko and collaborators 
\cite{Lvov-Nazarenko-arXiv-2010} and of Kozik and Svistunov
\cite{Kozik-Svistunov-JLTP-2009}, who predict $n_k \sim k^{-11/3}=k^{-3.7}$
and $n_k \sim k^{-17/5}=k^{-3.4}$ respectively, where their kelvon
occupation number $n_k$ is equivalent to our $A(k)$. 
The applicability of wave turbulence theory to the problem
is supported by our finding that the angle $\theta$
between the smoothed vortex line and the actual line is small and
has approximately a Gaussian nature, see Fig.~\ref{fig:17}.

The regime which we have identified in our calculation is clearly a weak
turbulent regime, as demonstrated by the small angles between the
smoothed vortex filament and the actual filament - see Fig.~\ref{fig:17}(top).
We think that in a stronger regime larger twists would appear and induce
more self-reconnections than the few which we observe in Fig.~\ref{fig:20}.
In must be stressed that our numerical cutoff effectively 
short-circuits the formation of loops at scales smaller than $\delta$. 
The tendency of the vortex filaments to twist could be explored more in
detail by examining the torsion, but we refrain from this calculation due to
larger errors which would arise in computing third derivatives. A better
approach would be to model the cascade more microscopically
using the Gross-Pitaevskii equation, although the range of scales which
would be available is limited.

The possibility that the geometry of superfluid turbulence undergoes
a qualitative change at low temperatures, where the friction can be ignored,
 was first considered
by Tsubota and collaborators \cite{Tsubota-Araki-Nemirovskii-PRB-2000}.
They noticed that if a vortex tangle is computed in the absence of
the smoothing effect of the mutual friction, its appearance is "kinky".
Further work \cite{Kivotides-fractal} showed that, even in the
presence of a small amount of friction, the vortex tangle is a fractal
object. This result stimulated the study of the
relation between energy and length, and between
energy and fractal dimension
\cite{Maggioni1,Maggioni2,Sciacca}. 
The results which we present here show that, in its simplest form, 
the fracture nature of superfluid turbulence arises from
the Kelvin-waves cascade along individual filaments.  
The actual value of the fractal dimension depends on the
level of acoustic dissipation of kinetic energy, which in our Biot-Savart
simulations is modelled by the numerical dissipation via
the parameter $\delta$. Therefore, for decaying turbulence as in our
calculations, $c_D$ depends on time. The fact that we observe some
secondary reconnections
only in the evolution shown in Fig.~\ref{fig:20} means that probably
fractality is a general feature of the Kelvin-waves cascade, not the
result of a sequence of self-similar twisting and
secondary reconnections; it would be interesting
to pursue this aspect of the problem with more computing power and a smaller
$\delta$.
It is worth remarking that in classical fluid mechanics
the relation between turbulence
and fractals has been noticed for some time \cite{Mandelbrot-fractal,
Sreeni-fractal,
Fung-Vassilicos-fractal, Procaccia-Brandenburg-fractal,Vanyan-fractal}.

Finally we notice that there are aspects of what we have discussed which
relate to classical fluid mechanics, particularly to the problem of
the complexity of solutions of the Euler equation for an inviscid
incompressible fluid. In the absence of viscous dissipation, classical
vorticity can stretch greatly, and filamentary vortex structures may twist
wildly\cite{Chorin,Siggia}, 
unable to reconnect, and may even tend to a blow-up in finite 
time. But there are important differences between incompressible
Euler vortices and
superfluid vortices.  Superfluid vortices
reconnect, lose energy by radiating it away, and vortex volume is
not conserved (the superfluid vortex core radius
is fixed by quantum mechanical constraints on the rotation). 
Therefore Lagrangian methods for classical flows and for superfluid flows
differ in both scope and detail. 
For example, Schwarz's de-singularization of the Biot-Savart integral does not 
involve a time-dependent core size. Another example is the cutoff $\delta$,
which models sound dissipation and the reconnecting distance.
In our case, dissipation and reconnections are essential ingredients
of the problem, and the numerical method must model their existence.
In the classical case, dissipation and reconnections are not
part of the problem.

\section{Acknowledgements}

We thank W.F. Vinen, E. Kozik and V. L'vov 
for suggestions and discussions. We are greatly indebted to
R. H\"{a}nninen, who stimulated the development of a better
energy-conserving algorithm.

%%%%%%%%%%%%%%%%%%%%%%%%%%%%%%%%%%%%%%%%%%%%%%%%%%%%%%%%%%%%%%%%%%%%%%%%%

\clearpage

%%%%%%%%%%%%%%%%%%%%%%%%%%%%%%%%%%%%%%%%%%%%%%%%%%%%%%%%%%%%%%%%%%%%%%%
\clearpage
%%%%%%%%%%%%%%%%%%%%%%%%%%%%%%%%%%%%%%%%%%%%%%%%%%%%%%%%%%%%%%%%%%%%%%%
\clearpage
%%%%%%%%%%%%%%%%%%%%%%%%%%%%%%%%%%%%%%%%%%%%%%%%%%%%%%%%%%%%%%%%%%%%%
\clearpage
%%%%%%%%%%%%%%%%%%%%%%%%%%%%%%%%%%%%%%%%%%%%%%%%%%%%%%%%%%%%%%%%%%%%%
\clearpage

%%%%%%%%%%%%%%%%%%%%%%%%%%%%%%%%%%%%%%%%%%%%%%%%%%%%%%%%%%%%%%%%%%%%%
\begin{figure}
\begin{minipage}{8cm}
\includegraphics[width=0.95\textwidth]{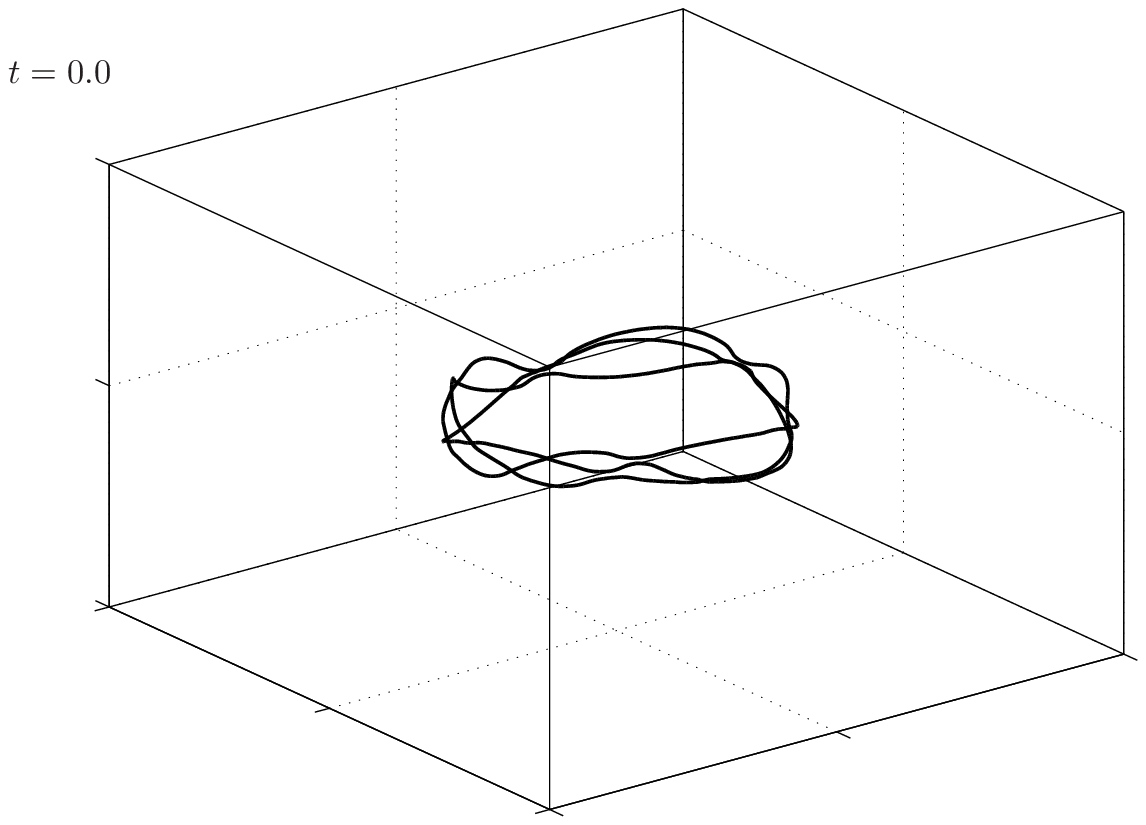}
\end{minipage}
\begin{minipage}{8cm}
\includegraphics[width=0.95\textwidth]{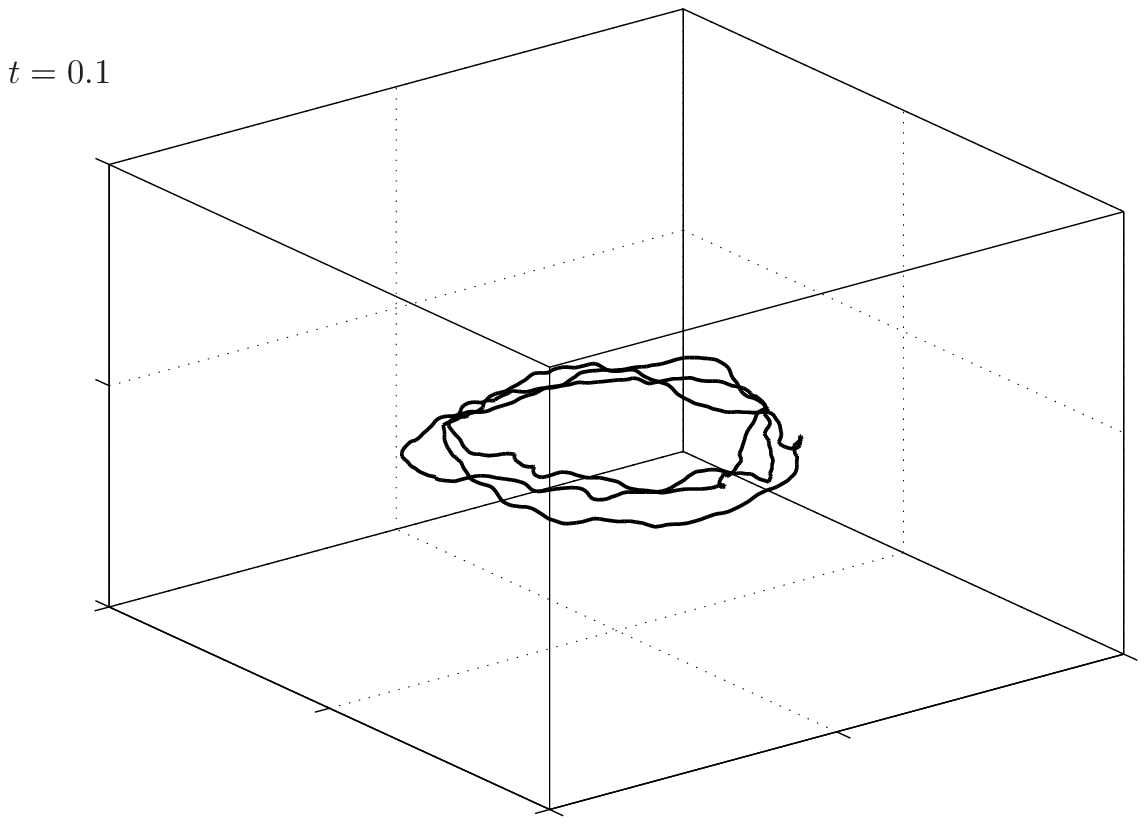}
\end{minipage}
\end{figure}
\clearpage
%%%%%%%%%%%%%%%%%%%%%%%%%%%%%%%%%%%%%%%%%%%%%%%%%%%%%%%%%%%%%%%%%%%%%

\clearpage

%%%%%%%%%%%%%%%%%%%%%%%%%%%%%%%%%%%%%%%%%%%%%%%%%%%%%%%%%%%%%%%%%%%%
\clearpage

%%%%%%%%%%%%%%%%%%%%%%%%%%%%%%%%%%%%%%%%%%%%%%%%%%%%%%%%%%%%%%%%%%%%%

\clearpage

%%%%%%%%%%%%%%%%%%%%%%%%%%%%%%%%%%%%%%%%%%%%%%%%%%%%%%%%%%%%%%%%%%%%%%%%%

\clearpage

%%%%%%%%%%%%%%%%%%%%%%%%%%%%%%%%%%%%%%%%%%%%%%%%%%%%%%%%%%%%%%%%%%%%%

\clearpage

%%%%%%%%%%%%%%%%%%%%%%%%%%%%%%%%%%%%%%%%%%%%%%%%%%%%%%%%%%%%%%%%%%%%%

\clearpage

%%%%%%%%%%%%%%%%%%%%%%%%%%%%%%%%%%%%%%%%%%%%%%%%%%%%%%%%%%%%%%%%%%%%%%%%
% no reconnections
\clearpage

%%%%%%%%%%%%%%%%%%%%%%%%%%%%%%%%%%%%%%%%%%%%%%%%%%%%%%%%%%%%%%%%%%%%%%%%

\clearpage

%%%%%%%%%%%%%%%%%%%%%%%%%%%%%%%%%%%%%%%%%%%%%%%%%%%%%%%%%%%%%%%%%%%%%
\clearpage

%%%%%%%%%%%%%%%%%%%%%%%%%%%%%%%%%%%%%%%%%%%%%%%%%%%%%%%%%%%%%%%%%%%%%
\clearpage

%%%%%%%%%%%%%%%%%%%%%%%%%%%%%%%%%%%%%%%%%%%%%%%%%%%%%%%%%%%%%%%%%%%%%
\clearpage

%%%%%%%%%%%%%%%%%%%%%%%%%%%%%%%%%%%%%%%%%%%%%%%%%%%%%%%%%%%%%%%%%%%%%%%%

\clearpage

%%%%%%%%%%%%%%%%%%%%%%%%%%%%%%%%%%%%%%%%%%%%%%%%%%%%%%%%%%%%%%%%%%%%%
\begin{figure}
\begin{minipage}{8cm}
\includegraphics[width=0.95\textwidth]{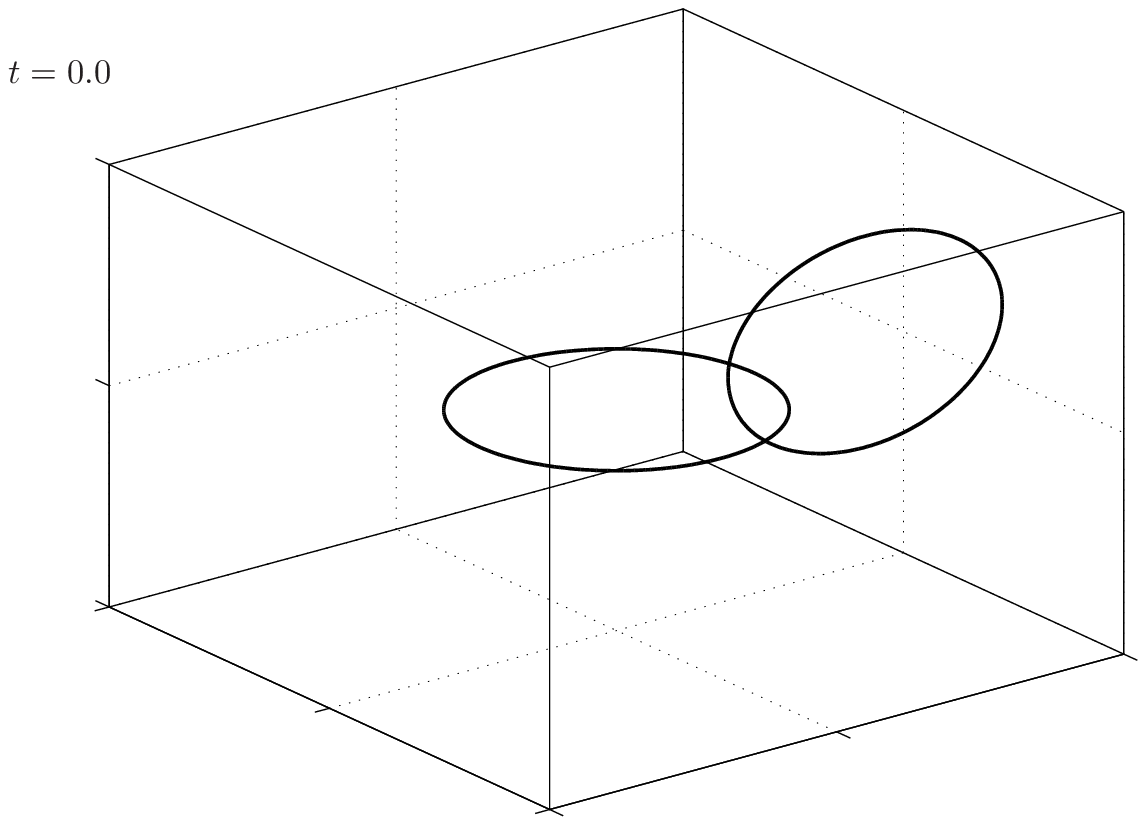}
\end{minipage}
\begin{minipage}{8cm}
\includegraphics[width=0.95\textwidth]{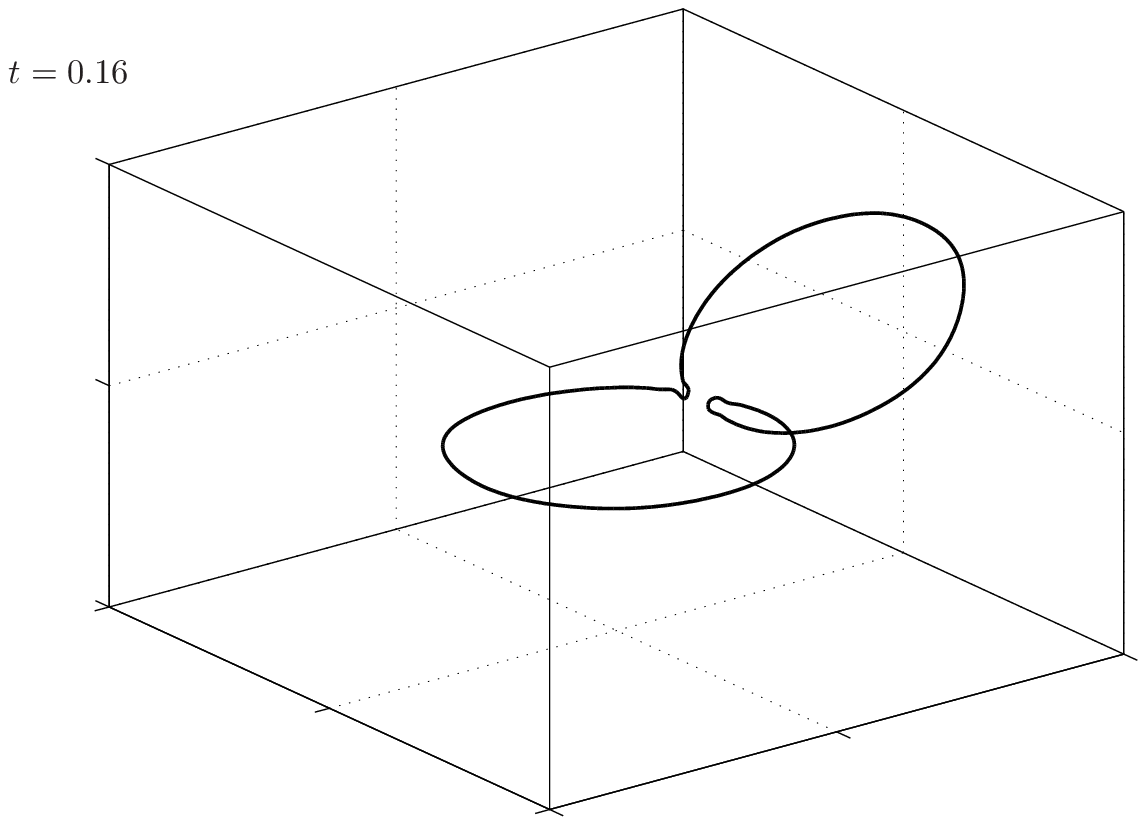}
\end{minipage}
\end{figure}
\clearpage

%%%%%%%%%%%%%%%%%%%%%%%%%%%%%%%%%%%%%%%%%%%%%%%%%%%%%%%%%%%%%%%%%%%%%
\clearpage

%%%%%%%%%%%%%%%%%%%%%%%%%%%%%%%%%%%%%%%%%%%%%%%%%%%%%%%%%%%%%%%%%%%%%%%

\begin{figure}
\begin{minipage}{8cm}
\includegraphics[width=0.95\textwidth]{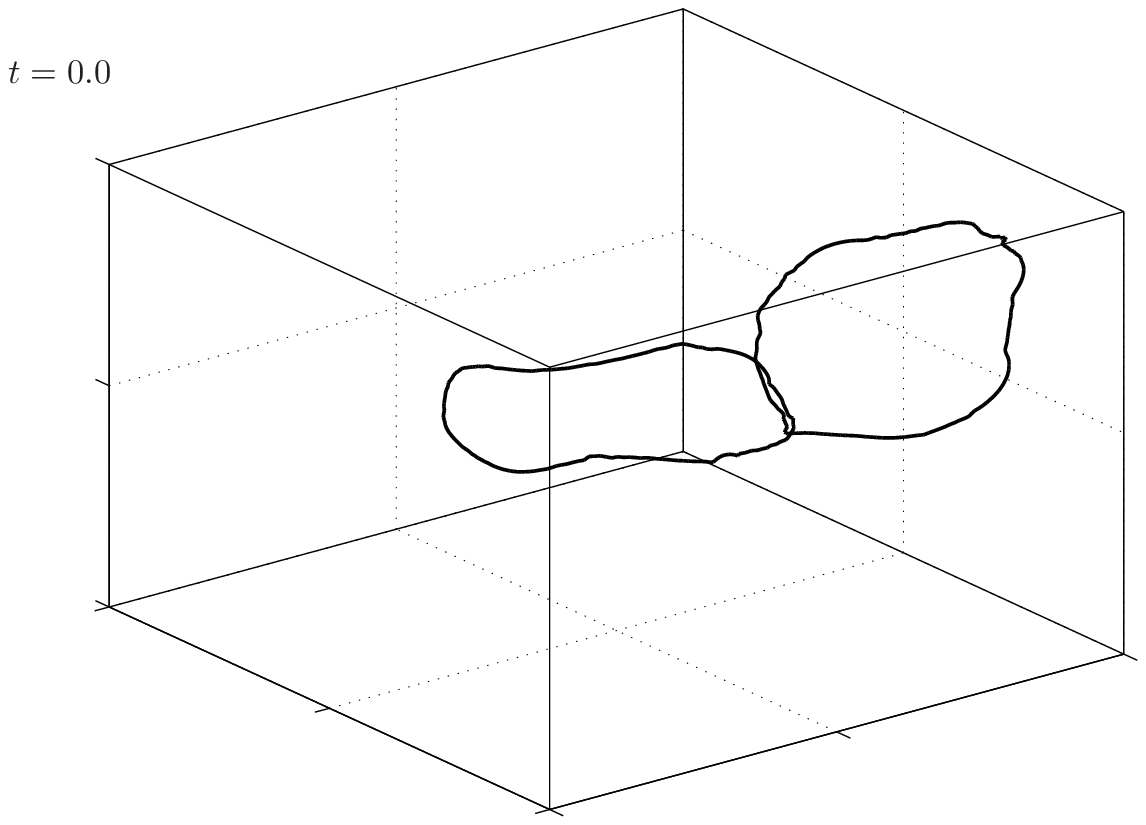}
\end{minipage}
\begin{minipage}{8cm}
\includegraphics[width=0.95\textwidth]{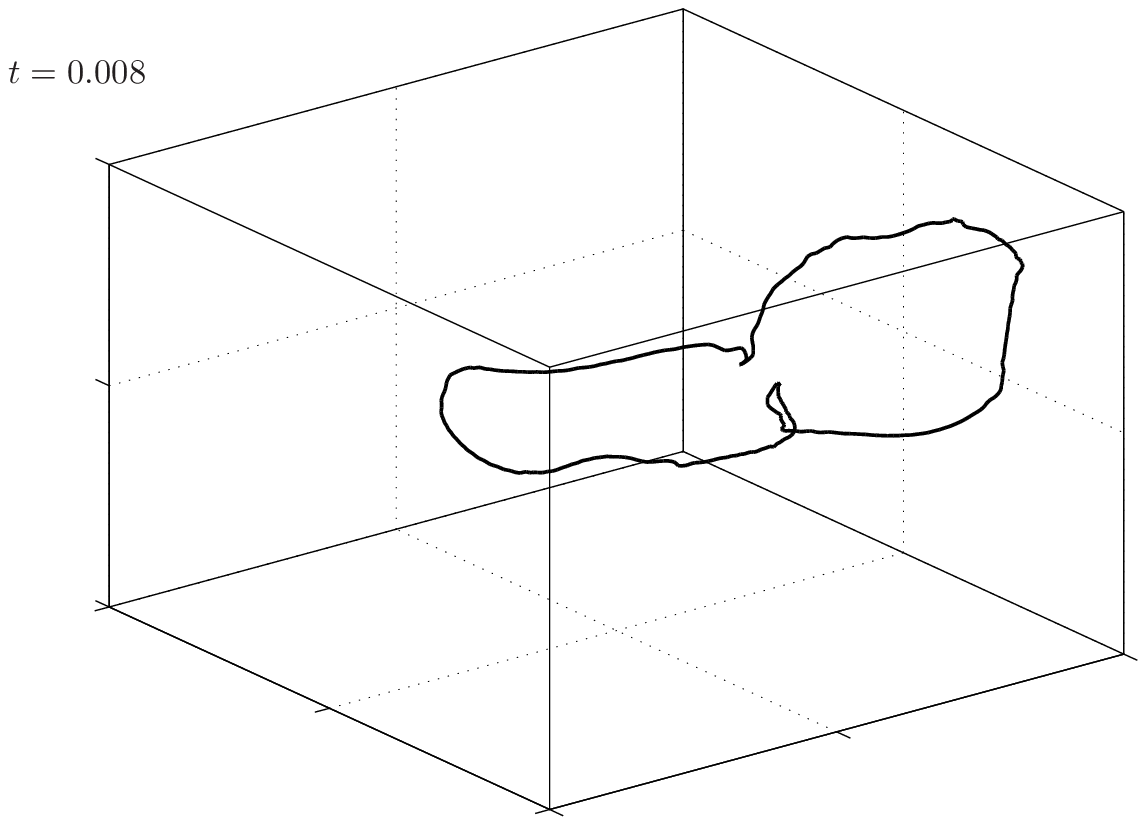}
\end{minipage}
\end{figure}
\clearpage
%%%%%%%%%%%%%%%%%%%%%%%%%%%%%%%%%%%%%%%%%%%%%%%%%%%%%%%%%%%%%%%%%%%%%%%

\clearpage

%%%%%%%%%%%%%%%%%%%%%%%%%%%%%%%%%%%%%%%%%%%%%%%%%%%%%%%%%%%%%%%%%%%%%

\clearpage
%%%%%%%%%%%%%%%%%%%%%%%%%%%%%%%%%%%%%%%%%%%%%%%%%%%%%%%%%%%%%%%%%%%%%
\end{document}